\documentclass[12pt,preprint]{aastex}
\usepackage{epsfig}
\usepackage{graphicx}
\usepackage{rotating}
\usepackage{booktabs}
\usepackage{multirow}
\usepackage{threeparttable}
\usepackage[page,title,titletoc,header]{appendix}

\newcommand{\kms}{$\mbox{km~s}^{-1}$}
\newcommand{\msun}{$M_{\odot}$}
\newcommand{\lsun}{$L_{\odot}$}
\newcommand{\jykms}{$\mbox{Jy~km~s}^{-1}$ }

\newcommand{\vlsr}{$V_{LSR}$}
\newcommand{\lbol}{$L_{bol}$}

\newcommand{\ml}{$\dot{M}_{loss}$}
\newcommand{\e}{$E_{k}$}
\newcommand{\p}{$P$}
\newcommand{\m}{$M_{outflow}$}

\newcommand{\deltav}{$\Delta$$v$}
\def\aj{AJ}%
\def\apj{ApJ}%
\def\apjl{ApJ}%
\def\apjs{ApJS}%
\def\aap{A\&A}%
\def\aaps{A\&AS}%
\def\azh{AZh}%
\def\mnras{MNRAS}%
\def\pasa{PASA}%
\def\nat{Nature}%
\shortauthors{Gan et al.}

\begin{document}

\title{A search for 95 GHz class I methanol masers in molecular outflows}
\author{Cong-Gui Gan\altaffilmark{1, 2, 3}, Xi Chen\altaffilmark{1, 2}, Zhi-Qiang, Shen\altaffilmark{1, 2}, Ye Xu\altaffilmark{4, 2}, Bing-Gang Ju\altaffilmark{4, 2}}
\altaffiltext{1}{Key Laboratory for Research in Galaxies and Cosmology, Shanghai Astronomical Observatory, Chinese Academy of Sciences, 80 Nandan RD, Shanghai, 200030, China}
\altaffiltext{2}{Key Laboratory of Radio Astronomy, Chinese Academy of Sciences, China}
\altaffiltext{3}{University of Chinese Academy of Sciences, No. 19A, Yuquan Road, Beijing, 100049, China}
\altaffiltext{4} {Purple Mountain Observatory, Chinese Academy of Sciences, Nanjing, 210008, China}
\email{cggan@shao.ac.cn}

\begin{abstract}

We have observed a sample of 288 molecular outflow sources including
123 high-mass and 165 low-mass sources to search for class I
methanol masers at 95 GHz transition and to investigate relationship
between outflow characteristics and class I methanol maser emission
with the PMO-13.7m radio telescope. Our survey detected 62
sources with 95 GHz methanol masers above 3$\sigma$
detection limit, which include 47 high-mass sources and 15
low-mass sources. Therefore the detection rate is 38\% for high-mass
outflow sources and 9\% for low-mass outflow sources, suggesting
that class I methanol maser is relatively easily excited in high-mass sources.
There are 37 newly detected 95 GHz methanol masers (including 27
high-mass and 10 low-mass sources), 19 of which are newly identified (i.e.
first identification) class I methanol masers (including 13
high-mass and 6 low-mass sources). Statistical analysis for the
distributions of maser detections with the outflow parameters reveals
that the maser detection efficiency increases with outflow
properties (e.g. mass, momentum, kinetic energy and mechanical
luminosity of outflows etc.). Systematic investigations of
relationships between the intrinsic luminosity of methanol maser and
the outflow properties (including mass, momentum, kinetic energy, bolometric luminosity and mass
loss rate of central stellar sources) indicate a positive
correlations. This further supports that class I methanol masers are
collisionally pumped and associated with shocks, where outflows
interact with the surrounding ambient medium.

\end{abstract}

\maketitle

\section{Introduction}

Methanol masers are widespread in our Galaxy, with more than 20
transitions in a wide frequency range from centimeter to
millimeter discovered to date \citep{cra05}. Their observed
connections with other star formation activities (e.g., infrared
dark clouds, millimetre and sub-millimetre dust continuum emissions
and ultracompact (UC) H{\sc ii} regions) made them one of the
most effective tools to investigate star forming regions
\citep[e.g.,][]{ell06}. Their trigonometric parallaxes provide
a direct and accurate measurement of distances to
star formation regions wherein methanol masers reside
\citep[e.g.,][]{xu06,ry10}. Their multiple frequency transitions
enable us to investigate the physical and chemical
conditions of star forming regions \citep[e.g.,][]{leu04,leu07,pur09}, and evolutionary
stages of star formation \citep[e.g.,][]{ell07,che11,che12}.

Methanol masers can be divided into two classes (class I and II)
according to the empirical classification on the basis of their
different exciting locations \citep{bat87,men91}. Class I methanol
masers are found usually offset ($\sim1'$, nearly ~1 pc at a
distance of 4 Kpc) from the presumed origin of excitation, and can
be further categorized to widespread class I methanol masers (e.g.,
44 and 95 GHz) and rare or weak class I methanol masers (e.g., 9.9
and 104 GHz) \citep{vor12}. The rare or weak masers trace stronger
shock regions which have higher temperatures and densities with
regard to widespread masers \citep{sob05, vor12}. In contrast,
class II methanol masers are often found to reside close to (within
1$\arcsec$) high-mass young stellar objects (YSOs) (e.g., Caswell et
al. 2010) and are frequently associated with UC H{\sc ii} regions,
infrared sources and OH masers. They also can be further categorized
to widespread (e.g., 6.7 and 12.2 GHz) and rare (e.g., 19.9, 23.1
and 37.7 GHz) class II methanol masers \citep{ell11,bar12}. Recently
\citet{ell11} found 37.7 GHz methanol masers (rare class II)
related with most luminous 6.7 and 12.2 GHz sources and thus
they suggested that the rare 37.7 GHz methanol masers are tracing more
evolved sources and arise prior to the cessation of widespread class
II methanol maser activity. The excitations of these two classes of
masers depend on two different pumping mechanisms: the pumping
mechanism of class I masers is dominated by collisions with molecular
hydrogen, whereas class II masers are pumped by external far-infrared
radiation \citep[e.g.,][]{cra92,vor99,vor05}. There is a competition
between the two mechanisms since strong radiation from a nearby infrared source
suppresses class I masers but strengthens class II masers
(see \citet[][]{vor05} for details). Surveys of class II methanol masers found that they
only exist in high-mass star forming regions \citep{min03,ell06,xu08};
while class I methanol masers have been detected not only in high-mass
star forming regions, but also in low-mass star forming regions
\citep{kal06,kal10}.

Many surveys of methanol masers have been carried out in last four decades.
The surveys of class II methanol masers (mainly at 6.7 GHz
transition) have detected nearly 900 sources to date
\citep[e.g.,][]{pes05,pan07,xu08,xu09,cyg09,gre09,gre10,gre12,cas10,cas11}.
While studies and surveys of class I methanol masers are rare
compared to class II methanol masers surveys. There are only a few single-dish
surveys \citep[e.g.,][]{has90,sly94,val00,ell05} as well as interferometric
searches \citep[e.g.,][]{kur04,cyg09}. However
class I masers have recently become the focus of more intensive
research
\citep[e.g.,][]{sar09,sar11,fon10,kal10,vor10a,vor10b,vor11,cha11,che11,che12,fis11,pih11}.
Some surveys have been carried out at the rare maser transitions,
e.g., at 9.9 GHz by \citet[]{vor10a} and 23.4 GHz by
\citet[]{vor11}. To date altogether $\sim$300 class I methanol maser
sources have been detected in our Galaxy \citep[see][for
details]{che11,che12}.

Earlier observations of class I methanol masers \citep[e.g.,][]{pla90,kur04}
had found that class I methanol masers located at the
interface regions between outflows and interstellar medium,
suggesting that locations associated with outflows may be one of the
best target sites for class I maser search. Statistical analysis
by \citet{che09} found that 67\% of outflow sources including
millimeter line molecular outflows (cataloged by Wu et al. 2004) and
EGOs, are associated with the class I methanol maser (at 95 GHz and 44 GHz)
within 1$'$. The EGOs are identified from $Spitzer$ Infra Red
Array Camera (IRAC) images in the 4.5 $\mu$m band, which is thought
to be a powerful outflow tracer and produced by shock-excited of
H$_2$ and CO \citep{cyg08}. A follow-up systematic survey towards a
nearly complete EGO sample (192 sources) with the Australia
Telescope National Facility (ATNF) Mopra 22-m radio telescope done
by \citet{che11} has detected 105 new 95 GHz class I methanol
masers, thus supporting a high detection rate (55\%) of
95 GHz methanol masers towards EGOs. In this paper we mainly focus on
95 GHz class I methanol maser searches in another outflow cataloged sample
included in the statistical analysis of \citet{che09} -- the outflow
sources identified from millimeter molecular spectral lines cataloged by
\citet{wu04} to check whether these millimeter molecular outflows have also
indeed a high detection rate of class I methanol maser as expected
in \citet{che09}. In addition, although the spatial distribution
relationship between class I methanol masers and outflows has been
investigated by a series of mapping observations
\citep[e.g.,][]{joh97,san03,san05}, most of these observations only
confirmed their spatial connections. The statistical studies of their
physical relationships (e.g., methanol luminosity and outflow
properties) are still absent. So it is also necessary to
perform a systematical search for class I methanol masers (e.g., 95 GHz) in
outflow sources to investigate the physical dependences between
outflows and the masers.

In this paper we report our result from the survey of 95 GHz class I
methanol maser toward the outflow sources selected from \citet{wu04}
outflow catalog. We describe our sample selection and observation in
$\S$ 2. In $\S$ 3, we present our results of class I methanol maser
detections. We discuss methanol maser detections
with outflow parameters and the relationships between outflow
parameters and maser luminosity in $\S$ 4. The conclusion is
summarized in $\S$ 5.

\section[]{OBSERVATION}

\subsection[]{Sample Selection}
Our sample sources are selected from \citet{wu04}, which cataloged a list
of molecular outflow sources ($\sim$400), identified from millimeters
molecular lines, along with their outflow parameters. These sources
are compiled mainly on the basis of mapping observations of CO at
low transitions ($J=1-0$ and $J = 2-1$), showing evidence of large
scale red- and blue-lobes. We choose the sources with Dec.$>$ -10
degrees, which can be accessible to the Purple Mountain Observatory
(PMO) 13.7-m telescope, and exclude sources with 95 GHz class I
methanol maser observed before the observing epoch, which included
in the statistical study of \citet{che09}. A total of 288
molecular outflow sources were selected for our survey. The sample
includes 123 high-mass and 165 low-mass sources according to their
available bolometric luminosity or outflow mass. \citet{wu04}
pointed out that the high-mass sources have bolometric luminosity of
larger than $10^3$ \lsun (for sources with bolometric luminosity
calculated) or outflow mass of larger than 3 \msun (for sources
without bolometric luminosity calculated), the others below these
limits are classified as low-mass sources. Their different mass
ranges are very helpful for comparing the 95 GHz class
I methanol maser detections and the relationships between outflow properties
and methanol masers. However, such
classification for high- and low-mass sources may not be reliable
for some cases. We will discuss this in Section 4.4.

\subsection[]{Observation and Data Reduction}
Single-point observations of 95 GHz class I methanol maser toward
the selected 288 sources were made in a period from 2010 June to 2010
September with the PMO 13.7-m telescope in Delingha, China. The rest
frequency of the observed $8_0 - 7_1 A^+$ transition is set to
95.169463 GHz. The half-power beamwidth of the telescope is about
$\sim$$55''$, and the pointing rms is better than $5'$ during the
observations. A cooled SIS receiver working in the 80$-$115 GHz band
was used and the system temperature was about 180$-$250 K, depending
on weather conditions. The spectra were recorded with an
Acousto-Optical Spectrometer (AOS) backend which has 1024 channels,
42 KHz for each channel and a total bandwidth of 42.7 MHz, resulting
in a velocity resolution of 0.13 \kms and total velocity coverage of
135 \kms. The observations were performed in position-switch mode
with off positions offset $15'$ from targeted point (no emission was
found in each off position). The aperture efficiency of the
telescope is 60\%, which implies that 1 K of antenna temperature
($T^{\star}_A$) corresponds to 31 Jy in flux density scale. The
observation was firstly carried out with an integrated time of 30
minutes for each source (achieving a typical rms noise level of 1.2
Jy), and then extend integration time (typical integration time is
60 minutes, resulting in a typical rms of 0.9 Jy) for the potential
weak sources.

The spectral data were reduced and analyzed with the GILDAS/CLASS
package. A first-order polynomial baseline subtraction was
performed for the majority of the observed sources, but for sources
with no good solutions from the first-order polynomial fits, we
carried out a second (or third)-order polynomial baseline
subtraction. After such a baseline removed, Hanning smoothing was
applied to obtain the spectra with a velocity resolution of 0.22 km
s$^{-1}$. Frequently the detected 95 GHz methanol spectra do not
exactly show a particularly Gaussian profile, possibly due to that
multiple maser features within a similar velocity range confuse the
spectra. However, each maser feature contributing to the complex
spectra often shows a single Gaussian profile. Thus, to characterize
the spectral characteristics of the total emission, we have performed
Gaussian fitting to each feature for each detected source.

\section[]{RESULT}

We have detected a total of 62 sources with 95 GHz methanol maser
emission flux above 3$\sigma$. A summary is given in Table~\ref{tab:sumdet}. The references which
were used to catalog newly-identified class I methanol maser and 95
GHz methanol maser are presented in Table~\ref{tab:highpar}~\&~\ref{tab:lowpar}. These
references include almost all the known class I methanol maser
surveys (including 36 GHz, 44 GHz and 95 GHz) to date. By cross-matching our detections with the
previously known-detected methanol masers from above references within
a spatial scale of 1$'$, we found 37 newly-detected 95 GHz
methanol maser sources. And among them 19 are newly identified as
class I methanol maser sources, i.e. the first identification of a
class I maser transition associated with these objects. This further
increases the sample of class I methanol maser, adding up to nearly
300 class I methanol maser sources \citep[see][]{che11,che12}.
Appendix~\ref{tab:undet} gives the undetected sources along with
their 1 rms noise, which have a range from 0.3 to 2.8 Jy, depending
on integration time and weather condition (a typical rms is 1 Jy).

\subsection[]{High-mass sources}

Our observations find 47 high-mass sources with 95 GHz
methanol masers above a detection limit of 3$\sigma$,
which include 27 newly-detected 95 GHz methanol masers and 13 of them are
newly-identified class I methanol masers. A list of the
detected 95 GHz methanol masers along with their Gaussian fitting
parameters in high-mass sources are presented in
Table~\ref{tab:highpar}, which includes three sub-tables:
(a) sources had 95 GHz class I methanol masers detected previously;
(b) sources detected at 95 GHz class I methanol masers in the first
time; (c) sources detected only at 95 GHz class I methanol masers
so far (i.e. Newly-identified class I methanol masers). We also listed information as to whether
the detected class I methanol masers are associated with class II
methanol masers or not in the table. Note that we only use the
catalog of 6.7 GHz class II methanol masers for which accurate
positions (better than 1$''$) have been published
\citep{xu09,cas09,cas11,gre10,gre12} in the cross-match. There are
13 sources which are found to be associated with class II methanol
masers within the measured figure of the outflow region in the 47
detected high-mass sources. Figure~\ref{fig:dethigh} shows the
detected 95 GHz methanol maser spectra and the fitted Gaussian
profiles with different color lines representing different fitted
components for these sources. The figure also shown in three
sub-figures according to whether sources had 95 GHz or other class I
methanol masers detected previously. The total integrated intensities of 95
GHz methanol maser range from 3.5 to 1070 \jykms with a mean of 47
\jykms for high-mass sources.

\subsection[]{Low-mass sources}
There are 15 sources which have 95 GHz methanol
maser emission detected with flux density above 3$\sigma$ in the low-mass
sample. Among them, 10 sources are newly-detected 95 GHz methanol
masers and 6 are newly identified as class I methanol
masers. The detected 95 GHz methanol masers along with their
Gaussian fitting parameters for low-mass sources are listed in
Table~\ref{tab:lowpar}, which is also subdivided into three sub-tables
according to whether sources had 95 GHz or other class I methanol
masers detected previously. We also listed information as to whether the detected 95 GHz
methanol maser sources are associated with class II methanol masers
or not in the table in the same approach as high-mass sources (see section 3.1). There
is only one source (G206.54-16.36, we will discuss this source in
Section 4.4) which is found to be associated with high accurate position
class II methanol masers in the 15 detected sources (see Table~\ref{tab:lowpar}). The
detected 95 GHz spectra and fitted Gaussian profiles for these
low-mass sources are presented in Figure~\ref{fig:detlow} and also
including three sub-figures as high-mass counterpart. The total
integrated intensities of 95 GHz methanol maser range from 1.3
\jykms to 45.5 \jykms with a mean of 12 \jykms for the low-mass
sources.

Therefore the maximal methanol intensity in high-mass sources is
nearly 24 times than that in low-mass sources, while the minimal
methanol intensity in high-mass source is 3 times that
in low-mass sources. The average intensity in high-mass sources is
nearly 4 times larger than that in low-mass sources.

\section[]{DISCUSSION}

\subsection[]{Detection Rates}

A total of 62 sources have been detected 95 GHz methanol masers
toward a sample of 288 outflow sources, giving a detection rate
of 22\% (62/288) for our survey. Out of 62 detections, 47 sources
are high-mass sources, thus a detection rate of 38\% (47/123) for
high-mass sources. The remaining 15 belong to low-mass sources, thus
a detection rate of 9\% (15/165) for low-mass sources.
However the actual detections/detection rates of methanol
masers may be affected by the following factors:  1) it can be clearly
seen from Figures 1 and 2 that only one single broad (and weak)
Gaussian profile was detected toward a number of sources (including
G70.29+1.60, G79.88+2.55, G105.37+9.84, G111.25-0.77, G173.58+2.44
and G213.70-12.60 in high mass source sample, and G65.78-2.61,
G183.72-3.66 and G208.77-19.24 in low mass source sample). We can
not determine whether they are thermal emission, or one or more maser
spectral features appending together from our current single dish
observations, although earlier high resolution observations to the
similar broad emission profiles showed that they are usually masers,
e.g, at 95 GHz \citep[]{vor06}, and 44 GHz \citep[]{cyg09,vor10b}; 2) it
should be noted that some detected methanol masers are located within
the PMO-13.7 m beam (e.g., high-mass sources
G111.53+0.76, G111.54+0.75 and G111.55+0.75 have angular separations
within 30$''$; low-mass sources G205.10-14.39 and G205.12-14.38 have
angular separations of $\sim40''$; low-mass source G206.56-16.36
and high-mass source G206.57-16.36 have angular separations of
$\sim25''$). If these nearby masers are only excited by the same one
source, the total detection rate would be 20\% (58/284; out of these
58 detections, 45 sources are high-mass sources, resulting in a
detection of 45/121=37\%; the remaining 13 sources belong to low-mass
group, thus a detection rate of 13/163=8\%). However, these nearby
detected masers show different spectral profiles, suggesting that
they may be excited by different driving sources; 3) the detection
rate may be affected by the possible extended spatial distribution of
class I methanol masers arising from sources with larger scale outflows (some
sources listed in \citet[][]{wu04} catalog show larger scale
outflows extending to several arc minutes). Therefore there is a possibility
that the detected maser emissions in one source may actually
originate from nearby sources with extended outflows to several
arc minutes along the line-of-sight. All above factors would
only be clarified with further higher resolution observations, but
these factors would not bring too much changes to the actual
detection rates. So we keep the methanol maser detection rates
derived from the current single dish observations in the following
subsequent discussions.

Interestingly, the detection rate of 22\% for the full observing
sample is in number nearly consistent with the finding of
\citet{val07}, which derived that 25\% of mm molecular line outflows
were associated with class I masers including 36 GHz, 44 GHz and 95
GHz within 2$'$ from a statistical analysis. However it is
significantly lower than that expected from the statistical analysis
of \citet{che09} for the \citet{wu04} outflow catalog. They have
analyzed 34 outflow sources from the \citet{wu04} catalog which have
been included in previous four class I methanol maser surveys
(including the 44 GHz transition by \citet{sly94} and
\citet{kur04}; and the 95 GHz transition by \citet{val00} and
\citet{ell05}). They found that 23 sources are associated with one
or both of the 95 and 44 GHz class I methanol masers within 1$'$,
thus the expected detection rate of class I masers in
\citet{wu04} outflow catalog is 67\% at this resolution. An actual
lower detection rate was achieved in the majority ($\sim$288/400) of
the cataloged outflow sources may be due to that previous
statistical study is subject to influences produced by the target
selection effects as followings: 1) only a small size sample which
includes 34 outflow sources was used in the statistical study; 2) in
previous four class I methanol maser surveys used in the study most
of target samples were pointed to UC H{\sc ii} regions, class II
masers and known class I maser sources; 3) the majority (29/34) of
sample used in previous analysis is high-mass sources which usually
show high detection rate of class I maser with regard to low-mass
sources (for example 38\% vs. 9\% in our observations, we will
discuss them in more details later); 4) previous statistical analysis
combined 44 and 95 GHz class I maser searches, and emission from the
44 GHz transition is generally 3 times stronger than that at 95 GHz
\citep{val00}, thus the search for methanol masers only at 95 GHz
transition is likely to have a lower detection rate than at 44 GHz or both 44
and 95 GHz transitions under comparable sensitivity. Combining these
it is not surprising for an overestimated detection rate of class I
methanol maser from previous statistical study.

Comparing the detection rate of 95 GHz class I methanol maser (55\%)
achieved toward nearly complete EGOs ($\sim$ 200) with the Mopra
telescope by \citet{che11}, we found that the actual detection rate
(of 22\%) of our 95 GHz class I masers in the full observing outflow
sample is also lower than that in the EGO sample. However, EGOs
trace a population of high-mass young stellar objects with ongoing
outflow activities. If only considering high-mass outflow sources in
our sample, the detection rate of 95 GHz class I maser of 38\% for
them is still smaller than that of 55\% for EGOs. The difference
between them is mainly due to the different detection sensitivities
in the two surveys: the typical detection sensitivity of $\sim$ 3 Jy
(3$\sigma$) in our observations is about two times that
of $\sim$ 1.6 Jy (3$\sigma$) in the \citet{che11} EGO surveys. To
check this, we re-examined the detected maser sources in the
\citet{che11} EGO surveys. When we excluded the sources (24 in
total) with maser peak flux density of less than $\sim$3 Jy from the
detected sources in the EGO survey, we found the detection rate is
81/192$=$42\%, which is nearly consistent with that achieved for
\citet{wu04} outflow catalog in this work at same sensitivity. This
also suggests that the EGOs have similar properties to outflow
sources traced by millimeter molecular lines from the view of their
correlation with 95 GHz class I methanol masers.

Note that our detection rate in low-mass sources is consistent with
that reported by \citet{kal10}, which detected four sources (NGC 1333I2A,
NGC 1333I4A, HH25MMS, and L1157) at 44 GHz, and one source (NGC 2023) at
36 GHz in a total of 44 low-mass outflow sources, resulting in a
detection rate of 11\%. Although the low-mass sources are closer to us than
high-mass sources on average, the detection rate and flux of
methanol maser in low-mass sources are lower than that in high-mass
counterparts on average in our observations. This suggests that the
high-mass sources have a higher outflow power than low-mass sources,
which could cause a higher collisional efficiency between
methanol molecule and surrounding clouds than low-mass sources (see
below).

\subsection[]{Detection rates with outflow parameters}

There are a series of outflow parameters including bolometric
luminosity of driving sources ($L_{bol}$), outflow mass ($M$),
momentum of outflows ($P$), kinetic energy of outflows ($E_k$),
force derived from outflow ($F$), mechanical luminosity of outflows
($L_m$), mass loss rate of central stellar sources
($\dot{M}_{loss}$) and dynamic time ($\tau$) associated with
outflows etc, presented in \citet{wu04}. We performed a series of
investigations on maser detections with each of the outflow
parameters. Histogram showing the detection rate of class I methanol
masers as a function of the bolometric luminosity of driving sources
($L_{bol}$), outflow mass ($M$), momentum of outflows ($P$), kinetic
energy of outflows ($E_k$), force derived from outflows ($F$),
mechanical luminosity of outflows ($L_m$), are presented in (a)--(f)
of Figure~\ref{fig:Det}, respectively. We adopt the mean value for
parameters which have been given more than one values in
\citet{wu04} in the analysis. The total sources and detected sources
are presented with different shapes in top panel of each diagram.
The bottom panel in each diagram denotes the corresponding detection
rate with the outflow properties. The corresponding detection rate
in each bin is represented by black dot and a low order polynomial
fit for the detection rates is marked with solid line (only fitted
for the data points with total observed source number larger than
5). All panels presented in Figure~\ref{fig:Det} show a clear
tendency that the detection rates of 95 GHz methanol maser increase
with the increment of outflow properties. With the increments of
outflow properties (e.g., outflow mass, momentum of outflows,
kinetic energy of outflows), much more materials would be
ejected in form of outflows, which would compress parent clouds and
increase methanol abundance, resulting in increasing collision
between methanol molecule and surrounding medium (mainly H$_2$) and
stimulating methanol molecule to higher energy levels. These
processes would ultimately cause a brighter maser excitation (we
will discuss this further in Section 4.3), making it more
easily to be detected in sources with higher outflow parameter values.

The methanol maser detection rates are related with outflow
properties from above analysis. We plot box plot to show
the significance of each outflow properties associated with methanol
maser presence, similar to the approach used in \citet{bre07,bre11}.
Figure~\ref{fig:boxp} shows result of box plot of the methanol maser
presence in consideration of outflow properties, which can be
divided into two categories of `n' and `y', corresponding to those
not related with methanol maser, and those related with methanol
maser. It can be clearly seen that the sources with
maser detected have a higher (larger) range of outflow properties
than the sources without maser detected. It suggests that these
outflow properties could play important roles in predicting methanol
maser presence. Methanol masers can be more easily detected in
sources with higher outflow properties than in those with lower
outflow properties relatively. This is consistent with above
detection rate analysis.

\subsection[]{Class I methanol maser emission with outflow properties}

The statistical studies of correlations between outflow properties
and class I methanol maser emission based on a large sample are
crucial for investigating the physical relationship between class I
methanol masers and outflows. It could be an important complement to
mapping observations \citep[e.g.,][]{pla90,kur04,vor06,cyg09}
-- the mapping observations were only made for limited size
sample, and they only present the spatial associations between
class I methanol masers and outflows at present. We have detected a
large number of 95 GHz methanol masers (62 in total) toward
molecular outflow sources in our observations. Most of the
detected sources were provided with the outflow properties in
\citet{wu04} catalog. All of them decide that our observations are
very suitable for such a statistical study. We performed a series
of analysis for the correlations between intrinsic luminosities of
detected methanol maser and outflow properties. The luminosities of
methanol maser can be calculated with $L =F_m\cdot4{\pi}\cdot d^2$,
where $L$ is the intrinsic luminosity of methanol maser, $F_m$ is
the total integrated intensity of 95 GHz methanol masers estimated
from Gaussian fitting, $d$ is the distance to outflow source. The
result indicates that there is significant correlation between
methanol maser intrinsic luminosity and outflow properties including
bolometric luminosity of central source ($L_{bol}$), outflow mass
($M$), momentum of outflows ($P$), mechanical luminosity of outflows
($E_k$). We show the log--log distributions of methanol maser
luminosity versus these four outflow properties in Figure 5 (a) --
(d). The green squares and red triangles in each panel of this
figure represent high-mass sources and low-mass sources,
respectively. The black solid line in each panel denotes the best
linear fit for each distribution. We give the best fitting results
in Table~\ref{tab:cor}. From this figure, we can clearly see that
most of high-mass sources reside at top right place in each panel,
with higher outflow properties and more luminous methanol masers,
whereas low-mass sources locate at the bottom left place, with lower
outflow properties and less luminous methanol masers. The
correlation coefficients for all these four relationships are larger
than 0.66, suggesting that strong correlations exist between 95 GHz
methanol maser luminosity and these outflow properties. This is
consistent with the theoretical expectation. The protostar ejects
materials in form of outflows, which squeeze clouds surrounding the
protostar. The generated shock propagating through high density
medium would stimulate methanol formation \citep{wir11} and enhance
methanol abundance \citep[e.g.,][]{gib98,gar02,vor10b}. These
effects combined would increase the collision efficiency of methanol
molecule with surrounding clouds and raise up the pumping efficiency
of class I methanol masers. As high-mass sources have higher outflow
power than low-mass sources, a brighter methanol maser would not be
unexpected to excite in high-mass sources. Our result for the
correlation between maser luminosity and bolometric luminosity of
outflow driving source is also comparable to the finding of
\citet{bae11}. They also demonstrated that there is a correlation
between bolometric luminosity of outflow driving source and
isotropic luminosity of only twelve 44 GHz methanol maser sources
detected in 180 intermediate-mass star forming regions, with a
correlation coefficient of 0.72.

The mass loss rate of central stellar sources directly reflect the
ejected materials from the central objects per unit time. So the
relationship between mass loss rate and class I methanol maser
luminosity is essential to interpret the dependence between methanol
masers and outflows. However, there are only a few sources in our
sample with mass loss rate of central stellar source estimates
presented by \citet{wu04}, so there shows a poor correlation between
them. We plot relationship between intrinsic luminosity of 95 GHz methanol
maser and mass loss rate of central stellar sources in  panel (e) of
Figure~\ref{fig:cor}. The best linear fitting result for this
dependence is also listed in Table~\ref{tab:cor}. Its correlation
coefficient is 0.33 due to the small size of the sample. But it also
shows the similar tendency to the other four outflow properties
discussed above, that the intrinsic luminosity of methanol maser
increases with the increment of mass loss rate of central stellar
sources (i.e., the flux of methanol maser is proportional to mass
loss rate of central stars in logarithm). This may further support
that class I methanol masers are collisionally excited, under which
with increment of outflow efficiency (e.g., outflow properties), the
phenomenon (e.g., shock) triggering methanol population inversion
is boosted up and hence more methanol excitations appear.

\subsection[]{Low-mass sources}

Studies of methanol maser in low-mass sources are an effective and
direct method in explaining the properties of methanol masers,
because the majority of detected low mass sources are closer to us
than high-mass sources. To date a total of 14 low-mass
sources have class I methanol masers detected including in one or
more transitions from 36 GHz, 44 GHz or 95 GHz. We list the previously known
class I methanol maser detections in low-mass sources in Table~\ref{tab:knmm}.
Among them, 8 sources have also been detected in our 95 GHz class I methanol maser
survey (see Table 5). Note that the previously detected source G205.11-14.38
is close to the two sources G205.10-14.39 and G205.12-14.38 detected in our
survey, with a separation of $\sim$20$''$ to each of the two sources respectively.
It means the previous detected source is located within the PMO-13.7 m beam of our detected two sources.
Thus we suggest that the two sources detected in our survey had class I methanol
masers detected previously. Our observations have found another 6 new
class I methanol masers at 95 GHz in low-mass sources, a significant
increase in the low-mass sample size. This also
confirms the existence of class I methanol maser in low-mass star
formation regions.

However we should note that the classifications of the high-mass and
low-mass sources on the basis of bolometric luminosity of the centeral
source or outflow mass proposed by \citet{wu04} may not be exact for
some cases. For example, the source G206.54-16.36 was classified as
low-mass sources due to its low outflow mass (0.04 M$_{\odot}$)
according to \citet{wu04}. But a 6.7 GHz class II methanol maser
which is exclusive tracer of high-mass star-formation has been
detected in this source, suggesting that it should be a high-mass
rather than low-mass star forming region. Therefore part of low-mass
sources classified by \citet{wu04} may not truly correspond to the
regions wherein only low-mass star forms. On the other hand, some
theories have proposed that high-mass star formation regions may
evolve from low-mass star formation regions \citep[e.g.,][]{arc07}.
If considering the above possible evolutionary effects, our observed
different mass type sources can be seen to locate at different
evolutionary stages. Our results shows that the detection rate of 95 GHz
class I methanol maser in high-mass sources are 4 times larger than
that in low-mass sources, meaning that the detection rate in more
evolved sources (i.e. high-mass regions) are also 4 times larger
than that in less evolved sources (i.e. low-mass regions). This
conclusion is consistent with \citet{fon10} which cataloged a total
of 88 sources and classified them into two groups including Low
sources and High sources according to their IRAS colours. The Low
sources are younger than the High sources according to their
criteria. Their result shows the detection rate of class I methanol
masers in High sources are nearly 3 times (2.9 times for 44 GHz and
3.3 times for 95 GHz methanol masers) than that in Low sources. This
also supports that more evolved sources are more easily detected
class I methanol maser than less evolved sources during evolutionary
stage of star formation. That is, with the source evolving from
low-mass to high-mass, the pumping efficiency also increases and
hence a brighter class I methanol maser is excited. Therefore as to
whether the class I masers detected in low-mass sources in our
observations are truly associated low-mass forming stars, we can
not completely exclude effects from the inaccurate mass-type
classifications and possible evolutionary effects from low-mass to
high-mass.

\section[]{SUMMARY}

A systematic survey of 95 GHz class I methanol masers was performed
towards 288 molecular outflow sources including 123 high- and 165
low-mass sources selected from Wu et al. (2004) outflow catalog with
the PMO-13.7 m telescope. We detected 62 sources with 95 GHz
class I maser above a detection limit of 3 $\sigma$, which include
47 high-mass sources and 15 low-mass sources. This suggests
that the detection rate of high-mass sources is 38\% and low-mass
sources is 9\%. The detection rate in high-mass sources is nearly 4
times that in low-mass sources, suggesting that 95 GHz class I
methanol masers are easily excited in high-mass sources. There are
37 newly detected 95 GHz methanol maser sources (including 27
high-mass sources and 10 low-mass sources), and 19 of them are
newly detected class I methanol maser sources (including 13
high-mass sources and 6 low-mass sources). This further increases
the number of the known class I methanol masers (adding on top of the
previous $\sim$300 class I maser sources) in our Galaxy. We performed
statistical analysis for the distribution of detection rates with
outflow properties. It shows a clear tendency that the distributions
of methanol maser detection rates increase with the increment of
outflow properties including outflow mass, momentum of outflows,
kinetic energy of outflows, bolometric luminosity of central source,
mechanical luminosity of outflows and force derived from outflows.
Analysis of the relationship between intrinsic luminosity of
methanol masers and outflow properties show that intrinsic luminosity
of methanol masers is logarithmically proportional to outflow
mass, momentum of outflows, kinetic
energy, and, bolometric luminosity and mass loss rate from
central stellar sources. This is in accord with the pumping
mechanism (collisionally excited) of class I methanol masers and
confirms the physical connections of methanol masers and outflows.

\section*{Acknowledgments}

We are grateful to the staff of Qinghai Station of Purple Mountain
Observatory for their help during the observations. This work is
partly supported by China Ministry of Science and Technology under
State Key Development Program for Basic Research (2012CB821800), the
National Natural Science Foundation of China (grants 10621303,
10625314, 10803017, 11121062, 10921063, 11073041, 11073054,
11133008, 11173046 and 11273043), the CAS/SAFEA International
Partnership Program for Creative Research Teams, and Key Laboratory
for Radio Astronomy, CAS.

\bibliographystyle{apj}

\newpage

\begin{table}
\centering
\caption{Summary of the 95 GHz class I methanol maser detected toward outflow sources}
\begin{tabular*}{0.8\textwidth}{lccc}
\hline
Summary &High-mass &Low-mass & Total \\
\hline
Outflow Targets & 123 & 165 & 288 \\
Detections & 47  & 15 & 62 \\
Detection rate & 38\% & 9\% & 22\% \\
$^a$Class I (new detections) & 27 (13) & 10 (6) & 37 (19) \\
\hline
\end{tabular*}\label{tab:sumdet}
\begin{tablenotes}
\item $^a$Number of class I methanol masers newly-identified at 95 GHz transition in the first time. The value in bracket present the number of newly-identified class I methanol masers in our work.
\end{tablenotes}
\end{table}

\begin{sidewaystable}
\caption[labelsep=space]{Detected 95 GHz methanol maser in high-mass sources}\label{tab:highpar}
\newsavebox{\tablebox}
\begin{lrbox}{\tablebox}
\begin{tabular}{lccccccccccccc}
\hline
Source &Other name&RA&DEC& {\vlsr} & {\deltav} & $P$ & $S$ & {$S_{int}$} & $RMS$ & \multicolumn{3}{c}{Class I} & Class II \\
\cline{11-13}
Name & & (J2000) & (J2000) & { ( \kms ) } & { ( \kms ) } & (Jy) & { ( \jykms ) } & { ( \jykms ) } & (Jy) & 36 GHz${}^a$ & 44 GHz${}^b$ & 95 GHz${}^c$& 6.7 GHz${}^d$ \\
(1)&(2)&(3)&(4)&(5)&(6)&(7)&(8)&(9)&(10)&(11)&(12)&(13)&(14)\\
\hline
\vspace{2mm}\\
\multicolumn{14}{c}{\Large (a) Sources had 95 GHz class I methanol masers detected previously.}\\
\hline
G10.84-2.59&GGD27&18:19:12.1&-20:47:26&13.15(0.04)&0.70(0.06)&72.98&54.12(8.56)&69.20&2.55&-&Y&Y&-\\
&&&&12.26(0.19)&0.92(0.75)&15.39&15.08(10.11)&&&&&&\\
G17.02-2.40&L379IRAS(2)&18:30:34.9&-15:14:38&14.80(0.11)&0.88(0.23)&3.94&3.67(0.87)&30.68&1.16&-&-&Y&-\\
&&&&18.49(0.21)&2.16(0.56)&6.43&14.75(3.41)&&&&&&\\
&&&&20.11(0.08)&1.09(0.20)&7.74&8.96(2.88)&&&&&&\\
&&&&22.65(0.27)&1.46(0.46)&2.13&3.30(1.02)&&&&&&\\
G19.88-0.54&18264-1152&18:29:14.7&-11:50:25&44.37(0.07)&0.82(0.14)&8.94&7.75(2.02)&59.19&0.85&-&-&Y&Y\\
&&&&43.42(0.19)&4.53(0.55)&4.41&21.27(2.54)&&&&&&\\
&&&&43.32(0.03)&1.14(0.08)&24.19&29.22(2.48)&&&&&&\\
&&&&40.98(0.05)&0.26(6.03)&3.40&0.95(0.40)&&&&&&\\
G25.65+1.05&18316-0602&18:34:20.8&-5:59:42&41.62(0.02)&0.54(0.04)&26.47&15.33(1.30)&58.54&2.44&-&Y&Y&-\\
&&&&43.85(0.04)&0.30(0.10)&8.60&2.77(0.87)&&&&&&\\
&&&&42.51(0.07)&0.37(0.13)&6.29&2.49(0.96)&&&&&&\\
&&&&42.95(0.21)&4.56(0.47)&7.83&37.95(3.66)&&&&&&\\
G35.20-0.74&G35.2-0.74&18:58:12.9&1:40:37&36.71(0.67)&2.56(1.00)&3.09&8.43(6.88)&50.31&1.30&-&Y&Y&Y\\
&&&&34.52(0.06)&1.92(0.26)&12.23&25.05(4.25)&&&&&&\\
&&&&33.35(0.63)&3.87(1.73)&4.08&16.83(3.19)&&&&&&\\
G43.17+0.00&W49&19:10:15.6&9:06:09&15.37(0.68)&4.68(1.06)&1.10&5.46(1.61)&8.41&0.53&Y&Y&Y&Y\\
&&&&14.31(0.08)&0.85(0.31)&2.22&1.99(0.84)&&&&&&\\
&&&&12.46(0.15)&0.71(0.30)&1.26&0.96(0.56)&&&&&&\\
G59.78+0.07&19410+2336(L)&19:43:11.3&23:44:06&22.98(0.29)&3.74(0.67)&1.97&7.82(1.64)&18.70&0.78&-&-&Y&-\\
&&&&22.31(0.05)&1.30(0.16)&6.33&8.74(1.58)&&&&&&\\
&&&&22.44(0.02)&0.35(0.06)&5.70&2.15(0.59)&&&&&&\\
G75.78+0.34&G75C&20:21:44.1&37:26:42&3.50(0.06)&1.08(0.14)&5.20&5.98(0.65)&14.70&0.79&Y&Y&Y&-\\
&&&&0.58(0.07)&0.93(0.15)&4.94&4.91(0.81)&&&&&&\\
&&&&-1.10(0.19)&1.55(0.44)&2.32&3.82(0.96)&&&&&&\\
G81.68+0.54&DR21&20:39:00.0&42:19:28&-3.68(0.01)&0.51(0.04)&20.47&11.00(0.87)&48.29&1.30&-&Y&Y&-\\
&&&&-2.00(0.08)&0.97(0.15)&5.14&5.32(1.27)&&&&&&\\
&&&&-3.12(0.10)&4.13(0.23)&7.28&31.96(2.17)&&&&&&\\
G81.88+0.78&W75-N&20:38:37.4&42:37:57&8.87(0.01)&0.37(0.03)&15.30&5.99(0.43)&62.22&1.06&Y&Y&Y&Y\\
&&&&11.92(0.09)&0.73(0.19)&2.54&1.98(0.59)&&&&&&\\
&&&&8.52(0.05)&4.25(0.13)&11.98&54.25(1.33)&&&&&&\\
G105.37+9.84&NGC7129 FIR&21:43:01.3&66:03:37&-8.53(0.14)&3.84(0.36)&1.61&6.58(0.50)&6.58&0.26&-&Y&Y&-\\
G106.80+5.31&S140&22:19:18.1&63:18:54&-7.15(0.13)&2.58(0.22)&2.95&8.09(0.81)&8.69&0.75&Y&Y&Y&-\\
&&&&-6.63(0.10)&0.43(0.19)&1.31&0.60(0.40)&&&&&&\\
G108.59+0.49&22506+5944&22:52:36.9&60:00:48&-51.11(0.26)&2.41(0.26)&2.86&7.32(0.34)&14.14&0.70&-&Y&Y&-\\
&&&&-49.63(0.26)&0.48(0.26)&1.17&0.59(0.34)&&&&&&\\
&&&&-51.90(0.26)&1.29(0.26)&2.18&2.99(0.34)&&&&&&\\
&&&&-46.45(0.26)&1.77(0.26)&1.72&3.23(0.34)&&&&&&\\
G111.53+0.76&NGC7538 IRAS11&23:13:44.7&61:26:54&-56.91(0.13)&1.13(0.13)&8.17&9.79(0.81)&89.82&1.74&-&Y&Y&-\\
&&&&-53.67(0.13)&3.03(0.13)&5.85&18.89(0.81)&&&&&&\\
&&&&-50.16(0.13)&1.38(0.13)&4.49&6.60(0.81)&&&&&&\\
&&&&-54.77(0.13)&2.72(0.13)&11.19&32.43(0.81)&&&&&&\\
&&&&-57.55(0.13)&2.37(0.13)&8.75&22.11(0.81)&&&&&&\\
G111.54+0.75&NGC7538 A&23:13:47.8&61:26:39&-50.80(0.18)&1.69(0.34)&2.19&3.94(0.84)&55.53&1.18&-&Y&Y&-\\
&&&&-53.24(0.03)&0.78(0.09)&8.62&7.11(1.30)&&&&&&\\
&&&&-54.95(0.08)&2.86(0.25)&12.04&36.62(2.98)&&&&&&\\
&&&&-57.06(0.06)&1.09(0.14)&6.80&7.85(1.64)&&&&&&\\
G111.54+0.78&NGC7538&23:13:44.7&61:28:10&-57.39(0.04)&1.00(0.18)&12.11&12.88(4.19)&31.76&1.25&Y&Y&Y&-\\
&&&&-57.02(0.20)&2.86(0.60)&6.20&18.88(3.97)&&&&&&\\
G173.49+2.44&05358+3543&5:39:10.6&35:45:19&-16.05(0.11)&1.26(0.27)&2.18&2.94(0.90)&13.84&0.50&Y&Y&Y&Y\\
&&&&-18.84(0.40)&5.58(0.85)&1.84&10.90(1.59)&&&\\
\end{tabular}
\end{lrbox}
\scalebox{0.6}{\usebox{\tablebox}}
\end{sidewaystable}

\begin{sidewaystable}
\addtocounter{table}{-1}
\caption{-- {\emph {continued}}}
\begin{lrbox}{\tablebox}
\begin{tabular}{lccccccccccccc}
\hline
Source &Other name&RA&DEC& {\vlsr} & {\deltav} & $P$ & $S$ & {$S_{int}$} & $RMS$ & \multicolumn{3}{c}{Class I} & Class II \\
\cline{11-13}
Name & & (J2000) & (J2000) & { ( \kms ) } & { ( \kms ) } & (Jy) & { ( \jykms ) } & { ( \jykms ) } & (Jy) & 36 GHz${}^a$ & 44 GHz${}^b$ & 95 GHz${}^c$& 6.7 GHz${}^d$ \\
(1)&(2)&(3)&(4)&(5)&(6)&(7)&(8)&(9)&(10)&(11)&(12)&(13)&(14)\\
\hline
G192.60-0.05&S255-IRS1&6:12:54.4&17:59:25&4.98(0.03)&0.30(0.08)&5.25&1.69(0.34)&19.35&1.10&-&Y&Y&Y\\
&&&&7.29(0.06)&0.94(0.17)&6.48&6.50(1.92)&&&&&&\\
&&&&8.45(0.50)&2.46(0.60)&2.94&7.70(2.02)&&&&&&\\
&&&&11.33(0.07)&0.73(0.16)&3.56&2.75(0.59)&&&&&&\\
&&&&9.62(0.14)&0.42(0.34)&1.57&0.71(0.78)&&&&&&\\
G208.99-19.38&ORION-A&5:35:14.5&-5:22:21&7.85(0.09)&1.54(0.16)&34.60&56.83(10.23)&1070.60&2.88&Y&Y&Y&Y\\
&&&&16.58(0.01)&12.39(0.34)&31.00&408.74(10.48)&&&&&&\\
&&&&6.53(0.06)&8.69(0.26)&34.62&320.42(9.55)&&&&&&\\
&&&&8.82(0.01)&3.62(0.09)&66.15&255.23(9.21)&&&&&&\\
&&&&8.60(0.02)&0.54(0.05)&51.66&29.39(4.74)&&&&&&\\
G209.01-19.41&ORION-S&5:35:12.4&-5:24:11&6.57(0.14)&3.05(0.16)&25.03&81.21(4.22)&143.69&1.92&Y&-&Y&Y\\
&&&&6.10(0.22)&2.10(0.38)&8.18&18.33(6.98)&&&&&&\\
&&&&8.35(0.24)&5.84(0.54)&7.10&44.15(5.18)&&&&&&\\
\hline
\vspace{2mm}\\
\multicolumn{14}{c}{\Large (b) Sources detected at 95 GHz class I methanol masers in the first time.}\\
\hline
G9.62+0.19&G9.62+0.19F&18:06:14.8&-20:31:39&5.82(0.58)&6.20(0.82)&1.50&9.90(7.75)&28.47&0.75&-&Y&-&Y\\
&&&&3.74(0.08)&2.55(0.21)&5.33&14.45(0.96)&&&&&&\\
&&&&0.41(0.20)&1.87(0.33)&2.07&4.12(0.87)&&&&&&\\
G45.07+0.13&G45.07+0.13&19:13:21.7&10:50:53&59.29(0.14)&0.95(0.41)&0.86&0.87(0.39)&5.79&0.25&-&Y&-&-\\
&&&&58.44(0.42)&6.36(1.01)&0.73&4.92(0.68)&&&\\
G53.03+0.12&19266+1745&19:28:53.9&17:51:56&5.05(0.18)&2.32(0.50)&1.47&3.62(0.62)&5.02&0.41&-&Y&-&-\\
&&&&6.71(0.09)&0.46(0.21)&1.20&0.59(0.31)&&&&&&\\
&&&&8.51(0.06)&0.46(0.11)&1.67&0.82(0.22)&&&&&&\\
G77.46+1.76&20188+3928&20:20:39.6&39:37:52&-2.87(0.19)&1.01(0.32)&1.43&1.53(0.50)&10.55&0.69&-&Y&-&-\\
&&&&-0.72(0.10)&0.85(0.20)&2.41&2.17(0.50)&&&&&&\\
&&&&1.72(0.10)&1.89(0.24)&3.39&6.84(0.74)&&&&&&\\
G78.12+3.63&20126+4104&20:14:26.0&41:13:32&-0.59(0.14)&0.61(0.35)&1.38&0.90(0.47)&24.18&0.75&-&Y&-&-\\
&&&&-4.61(0.06)&0.67(0.18)&3.12&2.23(0.87)&&&&&&\\
&&&&-3.29(0.02)&0.37(0.05)&6.01&2.36(0.40)&&&&&&\\
&&&&-3.01(0.09)&2.47(0.21)&7.11&18.69(1.43)&&&&&&\\
G78.97+0.36&20293+3952&20:31:10.4&40:03:10&6.59(0.09)&1.45(0.25)&3.67&5.66(0.78)&11.17&1.04&-&Y&-&-\\
&&&&4.98(0.06)&0.84(0.14)&4.99&4.46(0.68)&&&&&&\\
&&&&3.63(0.24)&0.89(0.38)&1.12&1.05(0.53)&&&&&&\\
G110.09-0.07&23033+5951&23:05:25.0&60:08:12&-54.14(0.15)&0.96(0.25)&6.00&6.15(2.26)&18.57&1.19&-&Y&-&-\\
&&&&-54.55(0.03)&0.35(0.08)&7.34&2.72(1.40)&&&&&&\\
&&&&-53.81(0.47)&5.52(1.31)&1.65&9.70(1.86)&&&&&&\\
G111.24-1.24&23151+5912&23:17:21.4&59:28:49&-51.73(0.13)&0.38(0.62)&1.32&0.54(0.37)&5.76&0.58&-&Y&-&-\\
&&&&-52.74(0.08)&0.85(0.17)&2.86&2.58(0.43)&&&&&&\\
&&&&-55.64(0.03)&0.26(0.95)&2.37&0.66(0.25)&&&&&&\\
&&&&-54.73(0.07)&0.73(0.17)&2.56&1.98(0.37)&&&&&&\\
G111.55+0.75&NGC7538 IRAS9&23:13:53.8&61:27:09&-57.32(0.12)&1.10(0.14)&1.63&1.91(0.34)&14.33&0.35&-&Y&-&-\\
&&&&-55.55(0.16)&0.90(0.40)&1.11&1.06(0.64)&&&&&&\\
&&&&-55.62(0.17)&4.92(0.49)&2.17&11.36(0.87)&&&&&&\\
G121.30+0.66&00338+6312&0:36:47.2&63:29:02&-17.80(0.20)&2.74(0.48)&2.57&7.49(1.21)&8.95&0.85&-&Y&-&Y\\
&&&&-17.56(0.06)&0.41(0.28)&3.29&1.45(0.74)&&&&&&\\
G122.01-7.07&00420+5530&0:44:57.2&55:47:18&-51.10(0.00)&1.09(0.22)&1.36&1.58(0.31)&3.49&0.34&-&Y&-&-\\
&&&&-48.50(0.00)&1.56(0.30)&1.15&1.91(0.37)&&&&&\\
G173.72+2.69&S235B&5:40:52.5&35:41:26&-16.40(0.01)&0.95(0.03)&15.22&15.31(0.37)&17.65&0.61&-&Y&-&-\\
&&&&-17.37(0.03)&0.43(0.08)&3.24&1.48(0.28)&&&&&&\\
&&&&-18.43(0.08)&0.49(0.21)&1.66&0.87(0.28)&&&&&&\\
G174.20-0.08&AFGL5142&5:30:46.0&33:47:52&-0.38(0.15)&1.00(0.25)&3.33&3.56(0.96)&23.73&1.54&-&Y&-&-\\
&&&&-3.15(0.03)&0.88(0.20)&7.92&7.38(3.16)&&&&&&\\
&&&&-4.62(0.50)&2.18(1.04)&2.60&6.05(2.79)&&&&&&\\
&&&&-1.90(0.13)&1.27(0.33)&5.00&6.75(1.36)&&&&&&\\
G188.95+0.89&AFGL 5180&6:08:54.2&21:38:25&2.27(0.45)&2.98(1.12)&3.49&11.08(2.08)&23.01&1.35&Y&-&-&Y\\
&&&&3.89(0.49)&2.74(0.93)&4.09&11.93(2.23)&&&&&&\\
\end{tabular}
\end{lrbox}
\scalebox{0.6}{\usebox{\tablebox}}
\end{sidewaystable}

\begin{sidewaystable}
\addtocounter{table}{-1}
\caption{-- {\emph {continued}}}
\begin{lrbox}{\tablebox}
\begin{tabular}{lccccccccccccc}
\hline
Source &Other name&RA&DEC& {\vlsr} & {\deltav} & $P$ & $S$ & {$S_{int}$} & $RMS$ & \multicolumn{3}{c}{Class I} & Class II \\
\cline{11-13}
Name & & (J2000) & (J2000) & { ( \kms ) } & { ( \kms ) } & (Jy) & { ( \jykms ) } & { ( \jykms ) } & (Jy) & 36 GHz${}^a$ & 44 GHz${}^b$ & 95 GHz${}^c$& 6.7 GHz${}^d$ \\
(1)&(2)&(3)&(4)&(5)&(6)&(7)&(8)&(9)&(10)&(11)&(12)&(13)&(14)\\
\hline
\vspace{2mm}\\
\multicolumn{14}{c}{\Large (c) Sources detected only at 95 GHz class I methanol masers so far (newly-identified class I methanol masers).}\\
\hline
G37.55+0.20&18566+0408&18:59:10.1&4:12:14&85.65(0.20)&4.33(0.54)&2.00&9.22(0.90)&9.22&0.53&-&-&-&-\\
G51.68+0.72&19217+1651&19:23:58.8&16:57:37&2.66(0.28)&2.93(0.90)&1.30&4.04(0.96)&8.23&0.48&-&-&-&-\\
&&&&-0.23(0.21)&1.00(0.38)&1.17&1.24(0.50)&&&&&&\\
&&&&4.66(0.11)&0.48(0.23)&1.29&0.66(0.37)&&&&&&\\
&&&&5.59(0.18)&0.73(0.36)&0.93&0.73(0.37)&&&&&&\\
&&&&7.55(0.18)&1.29(0.30)&1.13&1.56(0.40)&&&&&&\\
G70.29+1.60&19598+3324&20:01:45.6&33:32:43&-24.63(0.54)&7.24(1.03)&0.66&5.07(0.66)&5.07&0.19&-&-&-&-\\
G79.87+1.18&20286+4105&20:30:28.4&41:15:48&-3.12(0.23)&3.80(0.43)&2.78&11.24(1.36)&15.03&0.65&-&-&-&-\\
&&&&-4.09(0.08)&0.76(0.25)&3.01&2.42(0.87)&&&&&&\\
&&&&-6.79(0.11)&0.71(0.28)&1.81&1.38(0.47)&&&&&&\\
G79.88+2.55&20227+4154&20:24:31.1&42:04:17&4.84(0.58)&10.41(1.52)&1.34&14.82(1.67)&14.82&0.64&-&-&-&-\\
G109.99-0.28&23032+5937&23:05:23.2&59:53:53&-51.99(0.30)&0.43(0.49)&0.74&0.34(0.84)&5.11&0.52&-&-&-&-\\
&&&&-50.55(0.19)&1.19(0.44)&1.07&1.36(0.43)&&&&&&\\
&&&&-54.13(0.29)&1.00(1.61)&0.95&1.01(1.02)&&&&&&\\
&&&&-52.71(0.15)&0.93(0.62)&2.44&2.41(1.46)&&&&&&\\
G111.25-0.77&23139+5939&23:16:09.4&59:55:23&-44.98(0.12)&2.67(0.27)&2.54&7.23(0.62)&7.23&0.40&-&-&-&-\\
G123.07-6.31&00494+5617&0:52:23.9&56:33:45&-30.73(0.13)&5.88(0.36)&5.43&34.00(1.61)&45.28&0.81&-&-&-&Y\\
&&&&-40.31(0.28)&4.35(0.57)&2.43&11.27(1.33)&&&&&&\\
G138.30+1.56&AFGL4029&3:01:32.7&60:29:12&-33.98(0.10)&0.85(0.19)&1.66&1.50(0.34)&6.50&0.51&-&-&-&-\\
&&&&-36.67(0.14)&1.25(0.34)&1.99&2.65(0.62)&&&&&&\\
&&&&-38.05(0.11)&0.86(0.22)&2.08&1.91(0.53)&&&&&&\\
&&&&-39.12(0.05)&0.27(0.73)&1.53&0.44(0.19)&&&&&&\\
G173.58+2.44&05361+3539&5:39:27.5&35:40:43&-16.79(0.14)&1.58(0.32)&2.18&3.68(0.64)&3.68&0.54&-&-&-&-\\
G206.57-16.36&NGC2024 FIR 6&5:41:45.5&-1:56:02&11.77(0.10)&0.74(0.18)&1.96&1.54(0.40)&7.84&0.70&-&-&-&-\\
&&&&10.26(0.13)&1.88(0.20)&2.30&4.61(0.03)&&&&&&\\
&&&&7.16(0.07)&0.72(0.16)&2.20&1.69(0.31)&&&&&&\\
G207.27-1.81&AFGL961&6:34:37.6&04:12:44&12.09(0.13)&0.84(0.38)&2.84&2.54(0.76)&5.39&0.74&-&-&-&-\\
&&&&14.10(0.14)&0.68(0.33)&2.03&1.48(0.59)&&&&&&\\
&&&&17.27(0.13)&0.62(0.26)&2.09&1.37(0.56)&&&\\
G213.70-12.60&MON R2&6:07:48.3&-06:22:54&10.14(0.17)&2.34(0.39)&1.88&4.68(0.67)&4.68&0.47&-&-&-&Y\\
\hline
\multicolumn{13}{l}{Columns (1) present the source name sorted by the galactic coordinates.}\\
\multicolumn{13}{l}{Columns (2) - (4) list other source name from \citet[]{wu04} and corresponding equatorial coordinates.}\\
\multicolumn{13}{l}{Columns (5) - (9) present Gaussian fitting parameters of detected methanol emission: the velocity at peak \vlsr, the line FWHM {\deltav}, the integrated intensity $S$, the total}\\
\multicolumn{13}{l}{\hspace{0.85in} integrated intensity $S_{int}$. Value in brackets is the fitting error.}\\
\multicolumn{13}{l}{Columns (10) list 1-${\sigma}_{rms}$ noise of observations.}\\
\multicolumn{13}{l}{Columns (11) - (14) list the 36 GHz, 44 GHz and 95 GHz class I and 6.7 GHz class II methanol maser associations: Y = Yes, N = No, "-" = No information.}\\
\multicolumn{13}{l}{References:}\\
\multicolumn{13}{l}{${}^a$  \citet{lie96}}\\
\multicolumn{13}{l}{${}^b$  \citet{val07,kur04,lar07,fon10,kal06,kal10,lit11,bae11}}\\
\multicolumn{13}{l}{${}^c$  \citet{val95,val07,kal94,kal06,lar99,fon10,che11,che12}}\\
\multicolumn{13}{l}{${}^d$  \citet{xu09,cas09,cas10,cas11,gre10,gre12}}\\
\end{tabular}
\end{lrbox}
\scalebox{0.6}{\usebox{\tablebox}}
\end{sidewaystable}

\begin{sidewaystable}
\caption{Detected 95 GHz methanol maser low-mass sources}\label{tab:lowpar}
\begin{lrbox}{\tablebox}
\begin{tabular}{lccccccccccccc}
\hline
Name &Other name&RA&DEC& {\vlsr} & {\deltav} & $P$ & $S$ & {$S_{int}$} & $RMS$ & \multicolumn{3}{c}{Class I} & Class II \\
\cline{11-13}
Name & & (J2000) & (J2000) & { ( \kms ) } & { ( \kms ) } & (Jy) & { ( \jykms ) } & { ( \jykms ) } & (Jy) & 36 GHz${}^a$ & 44 GHz${}^b$ & 95 GHz${}^c$& 6.7 GHz${}^d$ \\
(1)&(2)&(3)&(4)&(5)&(6)&(7)&(8)&(9)&(10)&(11)&(12)&(13)&(14)\\
\hline
\vspace{2mm}\\
\multicolumn{14}{c}{\Large (a) Sources had 95 GHz class I methanol masers detected previously.}\\
\hline
G99.98+4.17&IC1396-N&21:40:42.1&58:16:10&0.00(0.04)&0.39(0.13)&3.21&1.32(0.40)&13.75&0.94&-&Y&Y&-\\
&&&&-0.84(0.03)&0.47(0.07)&5.44&2.75(0.53)&&&&&&\\
&&&&-0.76(0.19)&3.61(0.41)&2.52&9.68(1.02)&&&&&&\\
G158.40-20.57&NGC1333/IRAS4A&3:29:10.5&31:13:32&7.18(0.08)&0.30(2.67)&6.86&2.19(0.90)&17.39&1.42&-&Y&Y&-\\
&&&&6.81(0.12)&1.92(0.28)&7.44&15.20(1.92)&&&&&&\\
G205.10-14.39&HH26IR&5:46:05.5&-00:14:17&10.42(0.06)&0.83(0.16)&3.69&3.25(0.53)&6.17&0.53&-&Y&Y&-\\
&&&&9.17(0.13)&0.90(0.29)&1.83&1.74(0.53)&&&&&&\\
&&&&12.14(0.16)&0.43(0.77)&2.57&1.18(0.35)&&&&&&\\
G205.11-14.11&NGC2071&5:47:04.1&0:21:42&8.43(0.19)&1.34(0.37)&3.69&5.26(1.58)&14.64&0.86&Y&-&Y&-\\
&&&&12.96(0.41)&1.58(0.83)&1.35&2.27(0.99)&&&&&&\\
&&&&10.22(0.16)&1.59(0.46)&4.20&7.10(1.77)&&&&&&\\
G206.54-16.36&NGC2024 FIR 4&5:41:43.5&-1:54:45&9.64(0.39)&1.50(0.47)&2.40&3.83(0.65)&8.75&1.22&Y&-&Y&Y\\
&&&&10.31(0.51)&0.72(0.77)&2.19&1.68(0.84)&&&&&&\\
&&&&10.86(0.29)&0.91(0.30)&3.36&3.24(0.99)&&&&&&\\
\hline
\vspace{2mm}\\
\multicolumn{14}{c}{\Large (b) Sources detected at 95 GHz class I methanol masers in the first time.}\\
\hline
G65.78-2.61&20050+2720&20:07:06.2&27:28:53&6.32(0.17)&2.76(0.47)&1.12&3.29(0.43)&3.29&0.46&-&Y&-&-\\
G110.48+1.48&23011+6126&23:03:13.0&61:42:26&-11.09(0.26)&1.26(0.26)&3.15&4.21(0.31)&15.87&0.69&-&Y&-&-\\
&&&&-9.46(0.26)&2.02(0.26)&1.45&3.13(0.31)&&&&&&\\
&&&&-13.82(0.26)&2.08(0.26)&2.18&4.81(0.31)&&&&&&\\
&&&&-17.17(0.26)&0.46(0.26)&2.01&0.98(0.31)&&&&&&\\
&&&&-18.53(0.26)&2.04(0.26)&1.26&2.74(0.31)&&&&&&\\
G119.80-6.03&00259+5625&0:28:44.8&56:42:07&-38.55(0.14)&2.20(0.38)&8.37&19.62(3.94)&45.54&1.37&-&Y&-&-\\
&&&&-35.66(0.30)&2.98(0.71)&5.27&16.70(4.15)&&&&&&\\
&&&&-31.95(0.15)&2.14(0.36)&4.05&9.23(1.46)&&&&&&\\
G205.12-14.38&HH25 MMS&5:46:07.5&-0:13:36&12.45(0.16)&1.18(0.30)&1.79&2.25(0.56)&11.30&1.00&-&Y&-&-\\
&&&&11.02(0.09)&0.45(0.18)&2.04&0.98(0.40)&&&&&&\\
&&&&7.89(0.31)&2.54(0.66)&1.53&4.14(0.93)&&&&&&\\
&&&&10.18(0.04)&0.65(0.11)&5.72&3.93(0.62)&&&&&&\\
\hline
\vspace{2mm}\\
\multicolumn{14}{c}{\Large (c) Sources detected only at 95 GHz class I methanol masers so far (newly-identified class I methanol masers).}\\
\hline
G75.79+0.33&G75E&20:21:47.1&37:26:30&-0.41(0.23)&3.83(0.60)&1.58&6.44(0.80)&8.75&0.41&-&-&-&-\\
&&&&3.82(0.10)&1.09(0.21)&1.99&2.31(0.41)&&&&&&\\
G87.06-4.19&L944 SMM1&21:17:43.8&43:18:47&0.06(0.17)&1.57(0.48)&1.44&2.39(0.62)&8.00&0.52&-&-&-&-\\
&&&&4.60(0.14)&1.83(0.30)&1.96&3.80(0.56)&&&&&&\\
&&&&2.12(0.08)&0.88(0.30)&1.93&1.80(0.47)&&&&&&\\
G183.72-3.66&GGD4&5:40:24.1&23:50:54&2.64(0.26)&2.79(0.67)&1.17&3.47(0.67)&3.47&0.41&-&-&-&-\\
G206.56-16.36&NGC2024 FIR 5&5:41:44.5&-1:55:38&11.20(0.00)&1.05(0.41)&2.44&2.73(1.60)&9.78&0.46&-&-&-&-\\
&&&&11.00(0.00)&4.11(1.38)&1.61&7.05(1.38)&&&&&&\\
G208.77-19.24&FIR 1&5:35:21.8&-05:07:37&10.90(0.18)&1.56(0.35)&0.75&1.25(0.27)&1.25&0.25&-&-&-&-\\
G208.97-19.37&ORION-KLN&5:35:15.5&-5:20:41&9.84(0.06)&1.39(0.17)&6.60&9.73(0.90)&11.79&0.82&-&-&-&-\\
&&&&11.52(0.16)&0.98(0.37)&1.99&2.06(0.74)&&&&&&\\
\hline
\multicolumn{13}{l}{Columns (1) present the source name sorted by the galactic coordinates.}\\
\multicolumn{13}{l}{Columns (2) - (4) list other source name from \citet[]{wu04} and corresponding equatorial coordinates.}\\
\multicolumn{13}{l}{Columns (5) - (9) present Gaussian fitting parameters of detected methanol emission: the velocity at peak \vlsr, the line FWHM {\deltav}, the integrated intensity $S$, the total}\\
\multicolumn{13}{l}{\hspace{0.85in} integrated intensity $S_{int}$. Value in brackets is the fitting error.}\\
\multicolumn{13}{l}{Columns (10) list 1-${\sigma}_{rms}$ noise of observations.}\\
\multicolumn{13}{l}{Columns (11) - (14) list the 36 GHz, 44 GHz and 95 GHz class I and 6.7 GHz class II methanol maser associations: Y = Yes, N = No, "-" = No information.}\\
\multicolumn{13}{l}{References:}\\
\multicolumn{13}{l}{${}^a$  \citet{lie96}}\\
\multicolumn{13}{l}{${}^b$  \citet{val07,kur04,lar07,fon10,kal06,kal10,lit11,bae11}}\\
\multicolumn{13}{l}{${}^c$  \citet{val95,val07,kal94,kal06,lar99,fon10,che11,che12}}\\
\multicolumn{13}{l}{${}^d$  \citet{xu09,cas09,cas10,cas11,gre10,gre12}}\\
\end{tabular}
\end{lrbox}
\scalebox{0.6}{\usebox{\tablebox}}
\end{sidewaystable}

\begin{table}
\caption{Linear fitting results of Log-Log relationships between methanol maser luminosity and outflow properties.}
\begin{tabular*}{0.8\textwidth}{cccc}
\hline
Outflow properties & Slope & Intercept & correlation coefficient \\
\hline
\lbol & 0.514(0.081) & -7.769(0.326) & 0.66 \\
\m & 0.581(0.074) & -6.361(0.110) & 0.73 \\
\p & 0.616(0.079) & -7.036(0.179) & 0.74 \\
\e & 0.448(0.072) & -6.270(0.123) & 0.67 \\
\ml & 0.334(0.229) & -6.089(0.260) & 0.33 \\
\hline
\end{tabular*}\label{tab:cor}
\end{table}

\clearpage
\begin{table}
\caption{A list of previously known class I methanol maser low-mass sources.}
\begin{lrbox}{\tablebox}
\begin{tabular}{ccccccc}
\hline
Source & Other name & R.A. & Dec. & Frequency (GHz) & References & Our name \\
(1)&(2)&(3)&(4)&(5)&(6)&(7)\\
\hline
G65.78-2.61&20050+2720&20:07:06.7&27:28:53& 44 & 9 &G65.78-2.61\\
G99.98+4.17&Mol 138&21:40:41.0&58:16:16& 44,95&5,6,7&G99.98+4.17\\
G102.64+15.78& L1157 B1 & 20:39:08.1 & 68:01:13 & 44,95 & 2,3 & \\
G102.65+15.80& L1157-mm & 20:39:06.2 & 68:02:15 & 95 & 1 &  \\
G110.48+1.48&IRAS 23011+6126&23:03:13.0&61:42:26& 44 &8&G110.48+1.48\\
G119.80-6.03& CB 3 &0:28:42.7&56:42:07&44&8&G119.80-6.03\\
G158.36-20.58& NGC 1333I2A & 3:29:01.0 & 31:14:20 & 44 & 3 & \\
G158.39-20.57& NGC 1333I4A & 3:29:10.3 & 31:03:13 & 44 & 3 & \\
G158.40-20.58& NGC 1333IRAS4A & 3:29:12.0 & 31:13:09 & 44,95 & 2 & G158.40-20.57 \\
G205.07-14.36& HH 25MMS & 5:46:08.0 & 0:16:26 & 44,95 & 2 & \\
G205.11-14.11&NGC 2071&5:47:04.1&0:21:42& 36,95 & 4,5 & G205.11-14.11 \\
G205.11-14.38& HH25MMS & 5:46:06.5 & -0:13:54 & 44,95 & 3 & G205.10-14.39, G205.12-14.38 \\
G206.54-16.36& NGC 2024 &5:41:42.9&-1:54:34& 36,95 &4,5&G206.54-16.36\\
G206.89-16.60& NGC2023 & 5:41:28.5 & -2:19:19  & 36 & 3 & \\
\hline
\hline
\end{tabular}\label{tab:knmm}
\end{lrbox}
\scalebox{0.75}{\usebox{\tablebox}}
\begin{tablenotes}
\item Column (1) list source name sorted by the galactic coordinates.
\item Columns (2)--(4) present other source name and corresponding equatorial coordinates from references.
\item Columns (5)--(6) present frequency of previous detections and corresponding references.
\item Column (7) list corresponding low-mass source names with 95 GHz methanol maser detections in this work.
\item References: (1) \citet{kal01}; (2) \citet{kal06}; (3)\citet{kal10}; (4)\citet{lie96}; (5)\citet{val95}; (6) \citet{kur04}; (7)\citet{val07}; (8)\citet{bae11}; (9)\citet{fon10}.
\end{tablenotes}
\end{table}

\begin{figure}
    \psfig{figure=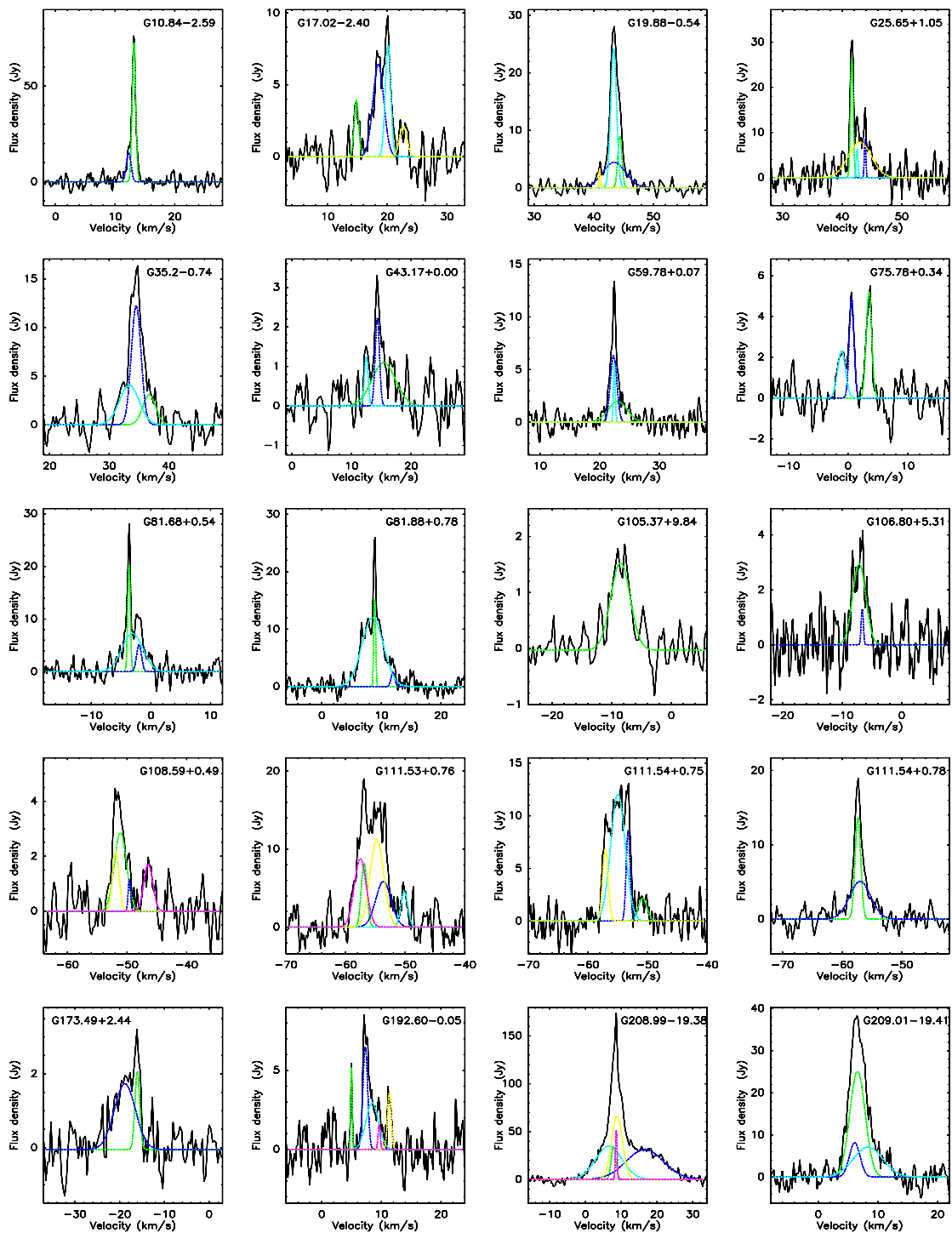,width=0.9\textwidth}\label{fig:ph11}
    \begin{center}
    \textbf{(a)}
    \end{center}
    \caption{A set of spectra of the detected 95 GHz methanol masers in high-mass
    sources: (a) sources had 95 GHz class I methanol masers detected previously; (b) sources detected at 95 GHz class I methanol masers in the first time; (c) sources detected only at 95 GHz class I methanol masers so far (newly-identified class I methanol masers). The dotted lines with different colors represent Gaussian fits to different
    velocity components (A color version of this figure is available in the online journal).}\label{fig:dethigh}
\end{figure}

\begin{figure}
    \psfig{figure=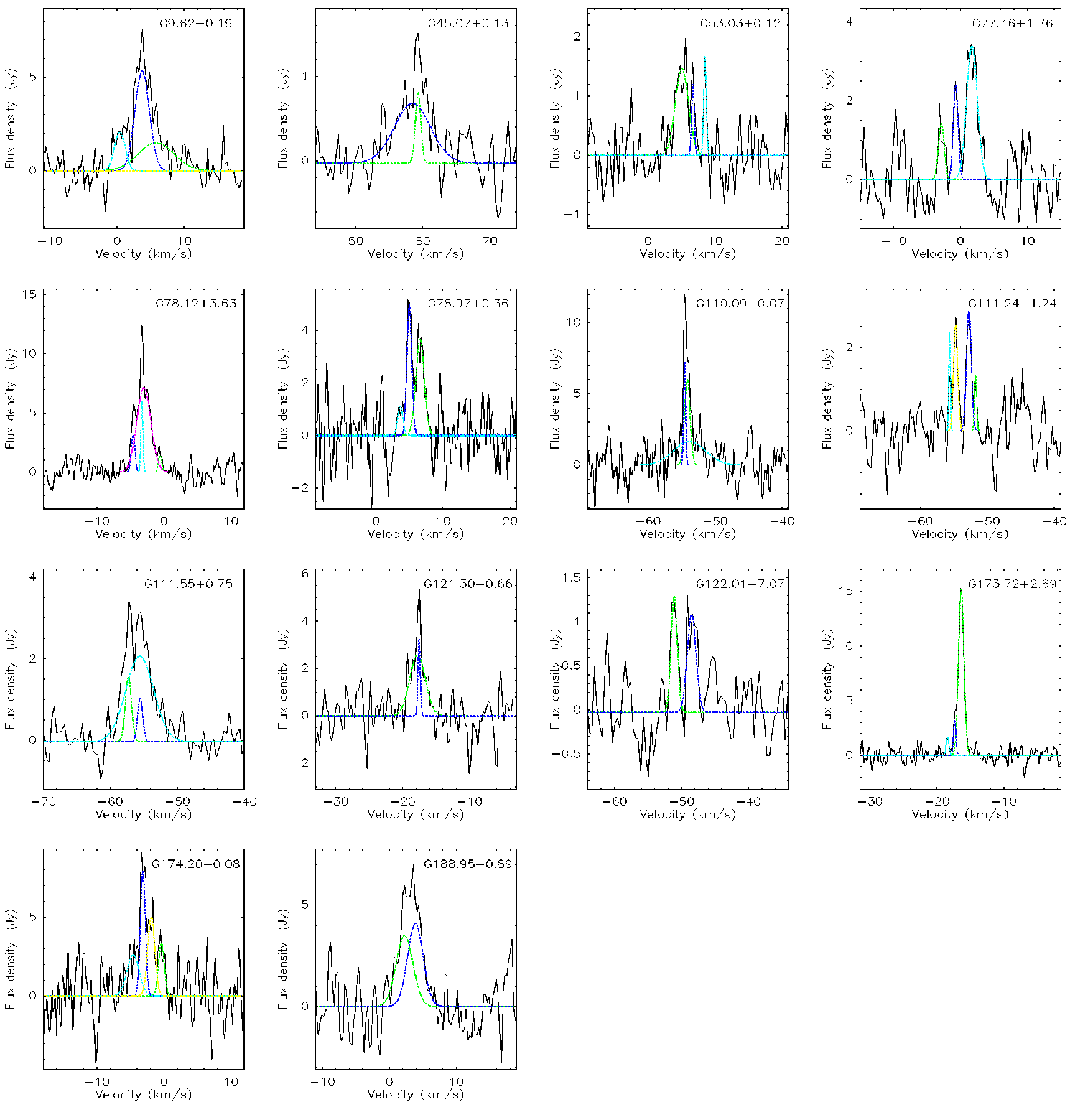,width=0.9\textwidth}\label{fig:ph22}
    \vspace{-5mm}
    \begin{center}
    \textbf{(b)}
    \end{center}
    \addtocounter{figure}{-1}
    \caption{-- {\emph {continued}}}
\end{figure}

\begin{figure}
    \psfig{figure=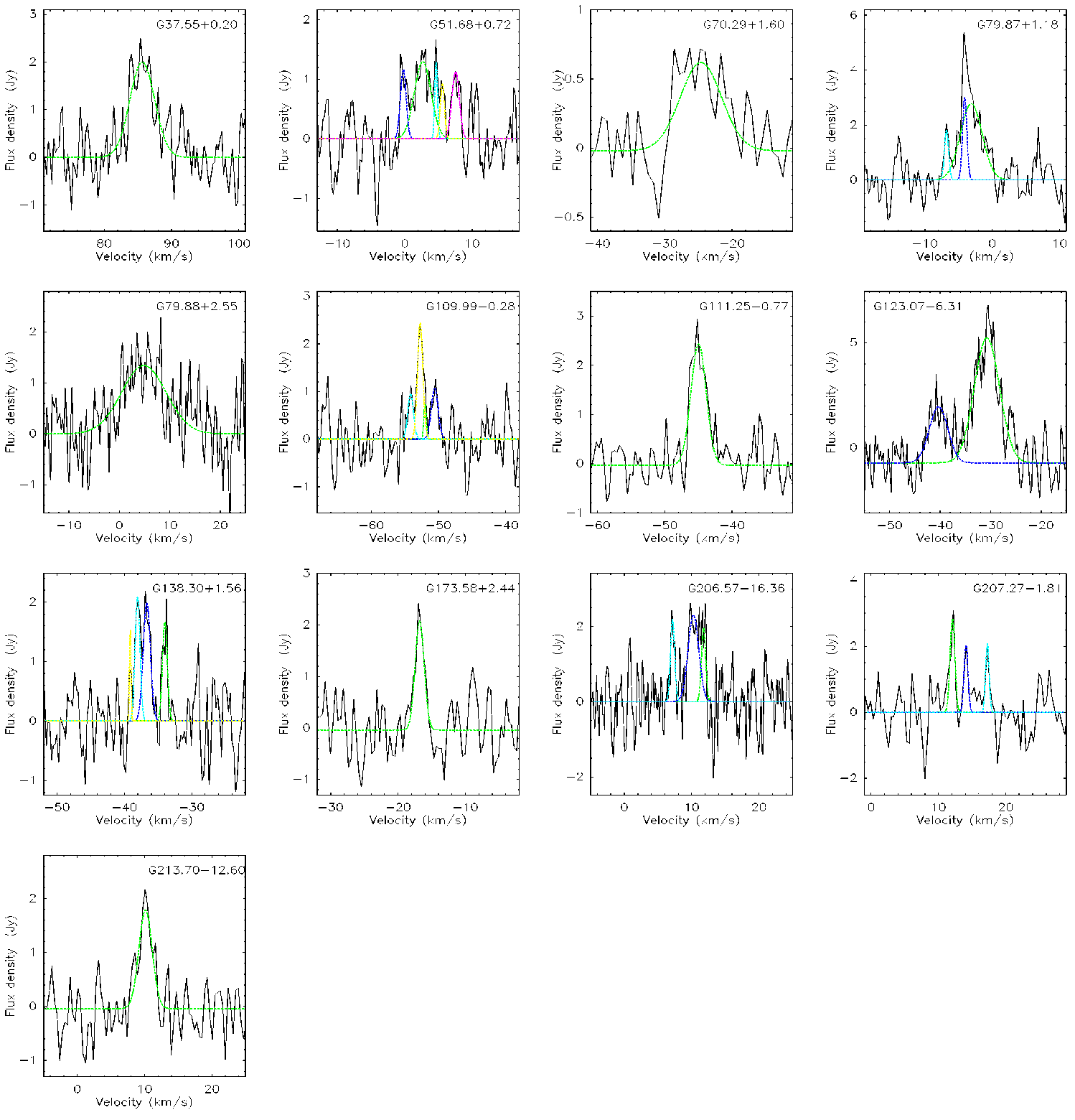,width=0.9\textwidth}\label{fig:ph33}
    \vspace{-5mm}
    \begin{center}
    \textbf{(c)}
    \end{center}
    \addtocounter{figure}{-1}
    \caption{-- {\emph {continued}}}
\end{figure}

\begin{figure}
    \psfig{figure=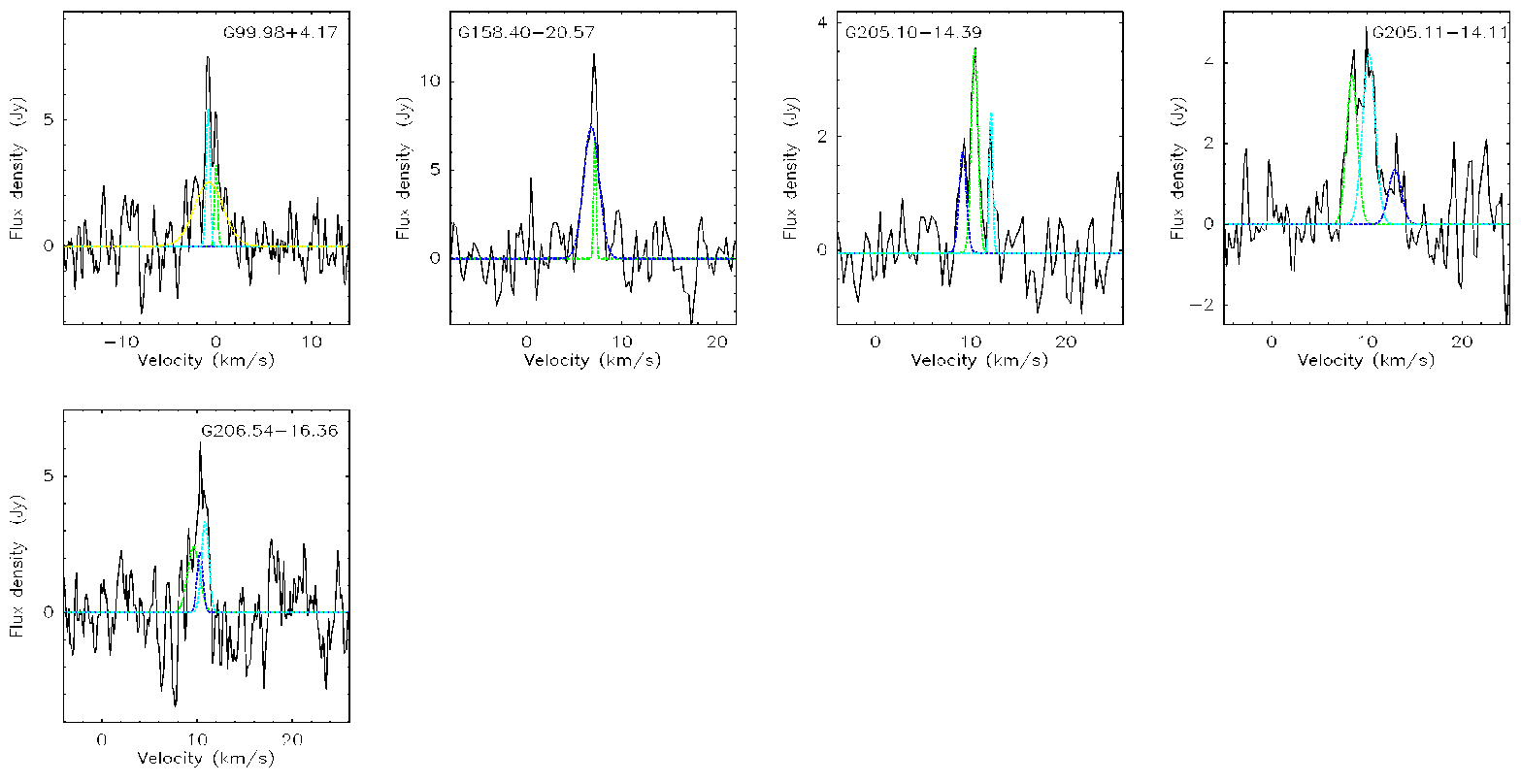,width=0.9\textwidth}
    \vspace{-5mm}
    \begin{center}
    \textbf{(a)}
    \end{center}
    \psfig{figure=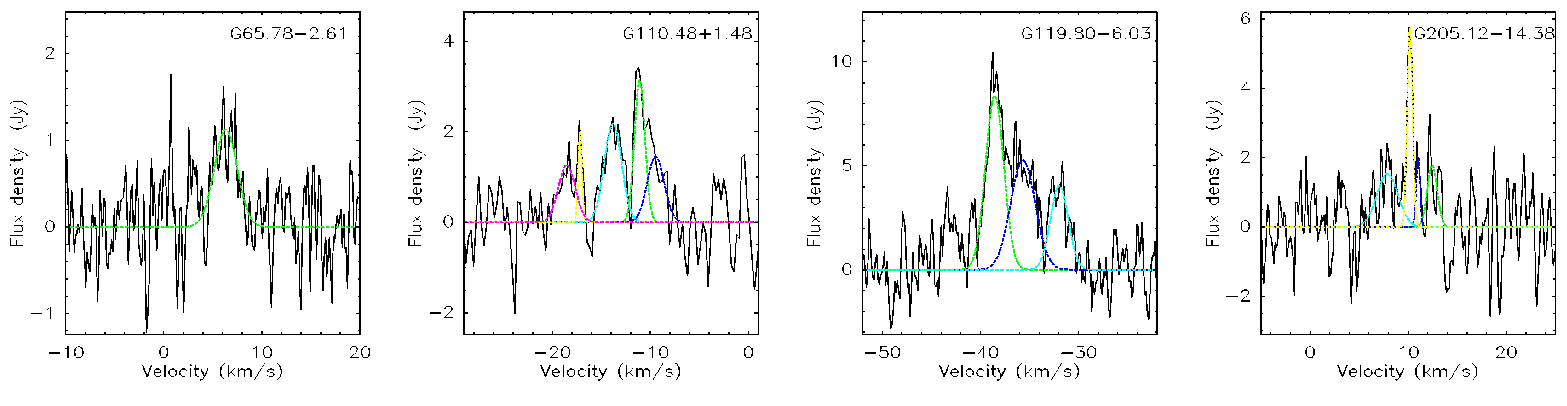,width=0.9\textwidth}
    \vspace{-5mm}
    \begin{center}
    \textbf{(b)}
    \end{center}
    \psfig{figure=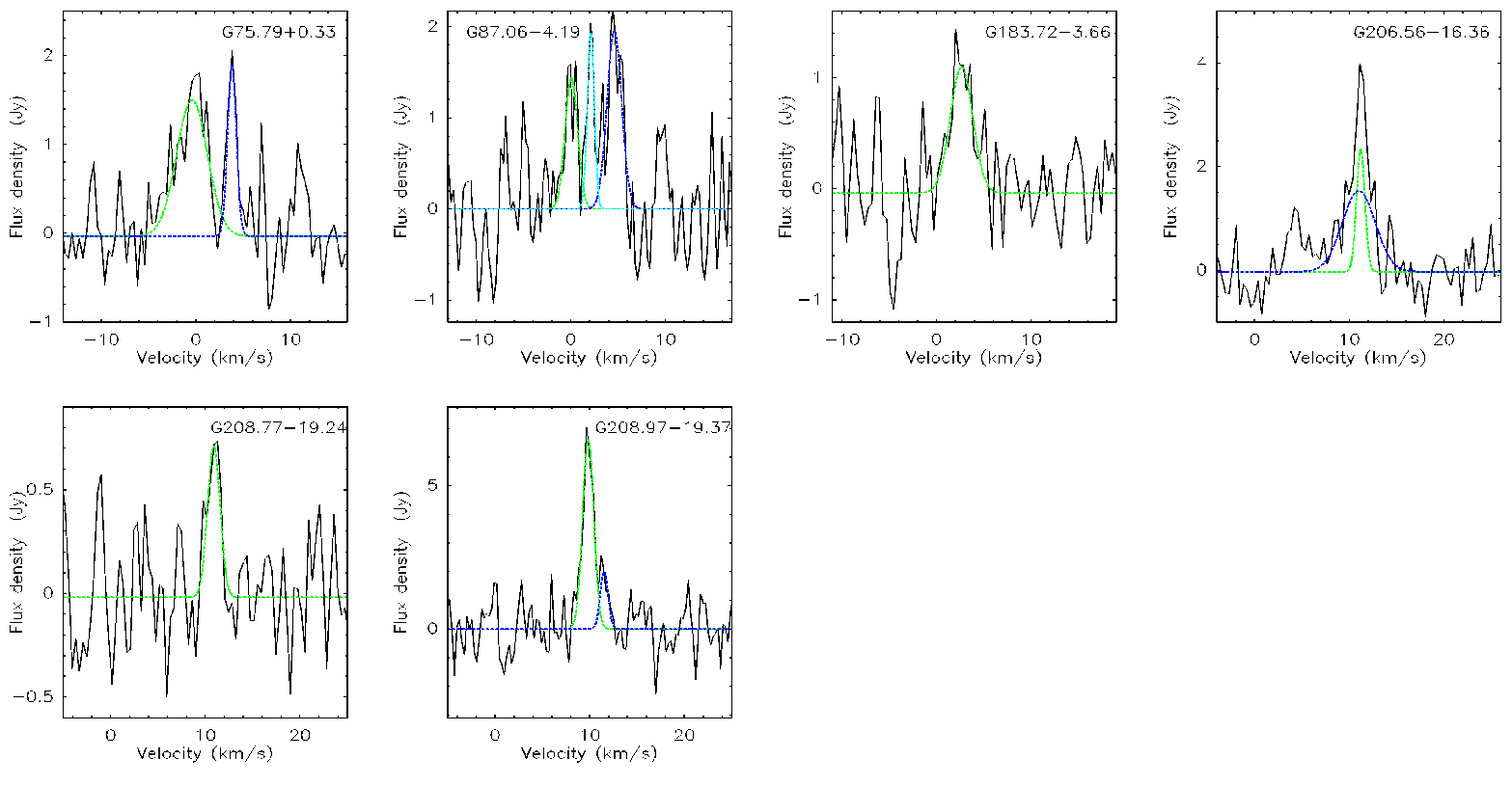,width=0.9\textwidth}
    \vspace{-5mm}
    \begin{center}
    \textbf{(c)}
    \end{center}
    \caption{A set of spectra of the detected 95 GHz methanol masers in low-mass
    sources: (a) sources had 95 GHz class I methanol masers detected previously; (b) sources detected at 95 GHz class I methanol masers in the first time; (c) sources detected only at 95 GHz class I methanol masers so far (newly-identified class I methanol masers). The dotted lines with different colors represent Gaussian fits to different velocity components (A color version of this figure is available in
    the online journal).\label{fig:detlow}}
\end{figure}

\begin{figure}
    \psfig{figure=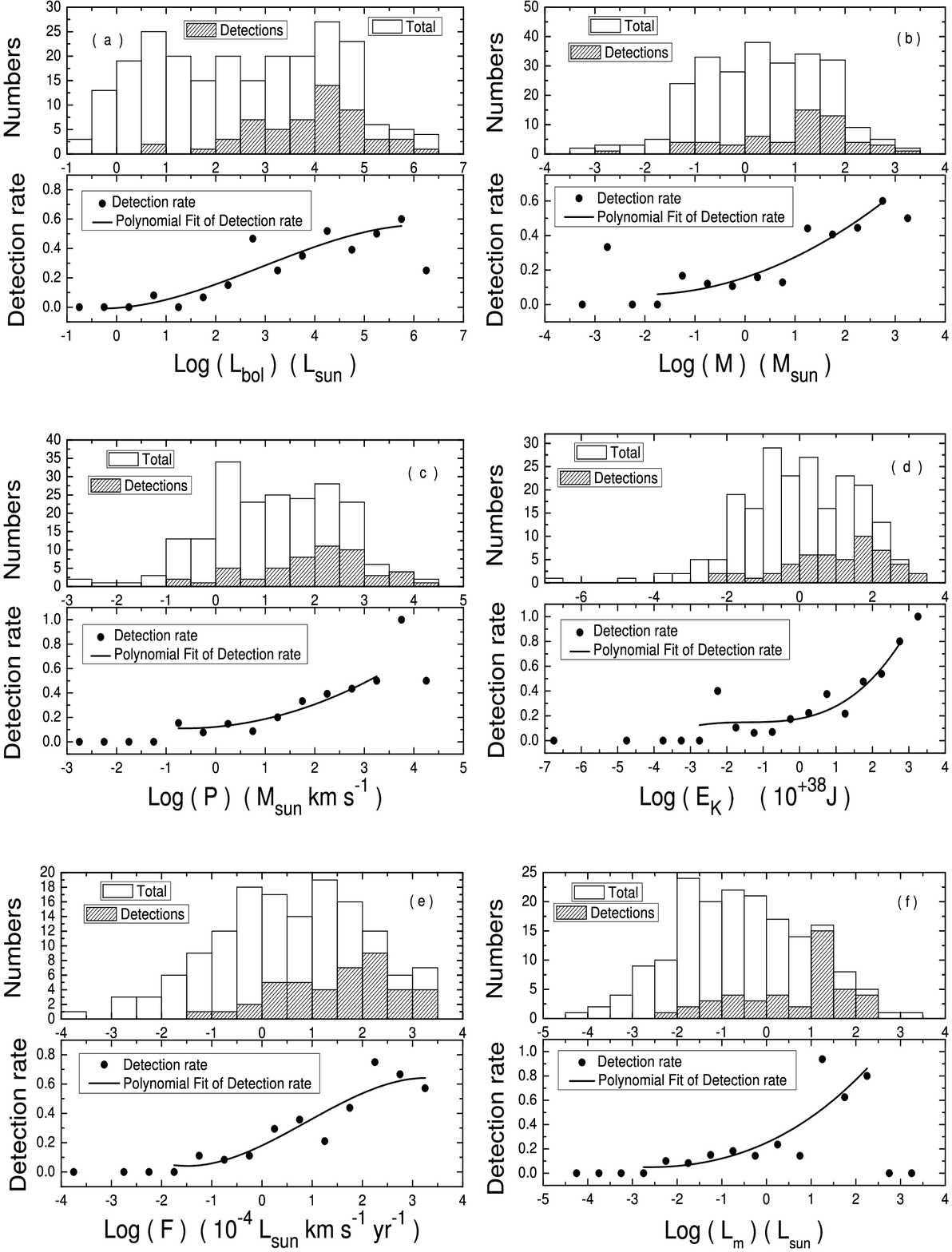,width=0.9\textwidth}
    \caption{The detection statistics of outflow properties including bolometric luminosity of central source (a),
    outflow mass (b), momentum of outflows (c), kinetic energy of outflows (d), driving force of outflows (e) and
    mechanical luminosity of outflows (f). The boxes shown with white boxes and shaded boxes in the top of each diagram represent total observed and detected sources. The dots in the bottom of each diagram represent the corresponding detection rates in each bin and solid line marks the lower-order polynomial fits for the detection rates. Note that the polynomial fits are only for the data points where the total source number is larger than 5.\label{fig:Det}}
\end{figure}

\begin{figure}
    \psfig{figure=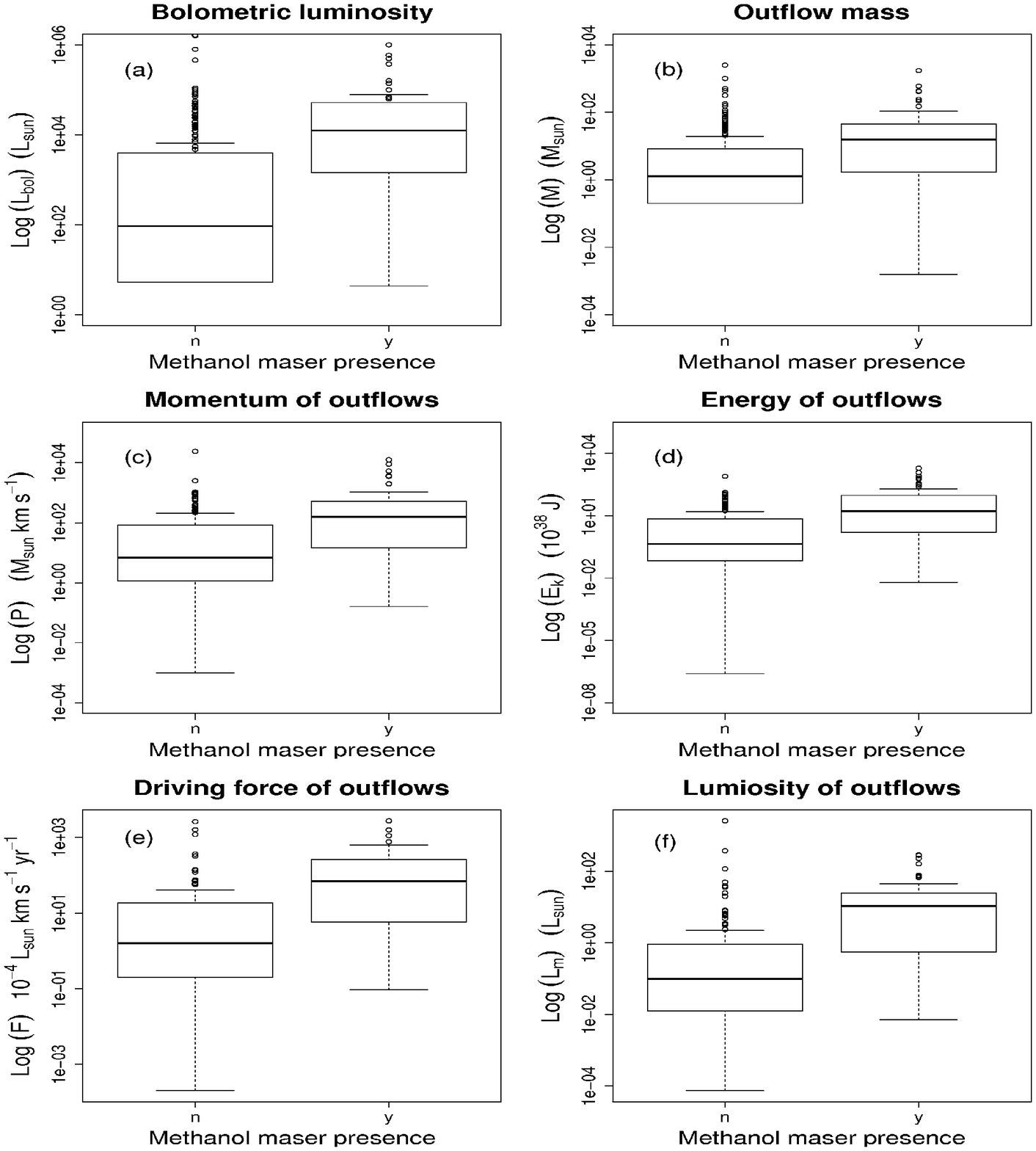,width=0.9\textwidth}
\caption{Statistical box plots of the molecular outflow properties presented in
the categories of yes and no, according to the presence of methanol
maser emission. There is distinct detection difference in outflow
properties including bolometric luminosity of central source (a), outflow mass
(b), momentum of outflows (c), kinetic energy of outflows
(d), driving force of outflows (e), mechanical luminosity of
outflows (f). The box contains data from 25\% to 75\%, the
line within each box represents the median of the data.
The vertical lines from the top of the box and from the bottom to the box
represent from 75\% to maximum value and from 25\% to minimum
value,respectively. The open circles represent the outliers, which were excluded from the statistics.
\label{fig:boxp}}
\end{figure}

\begin{figure*}
    \psfig{figure=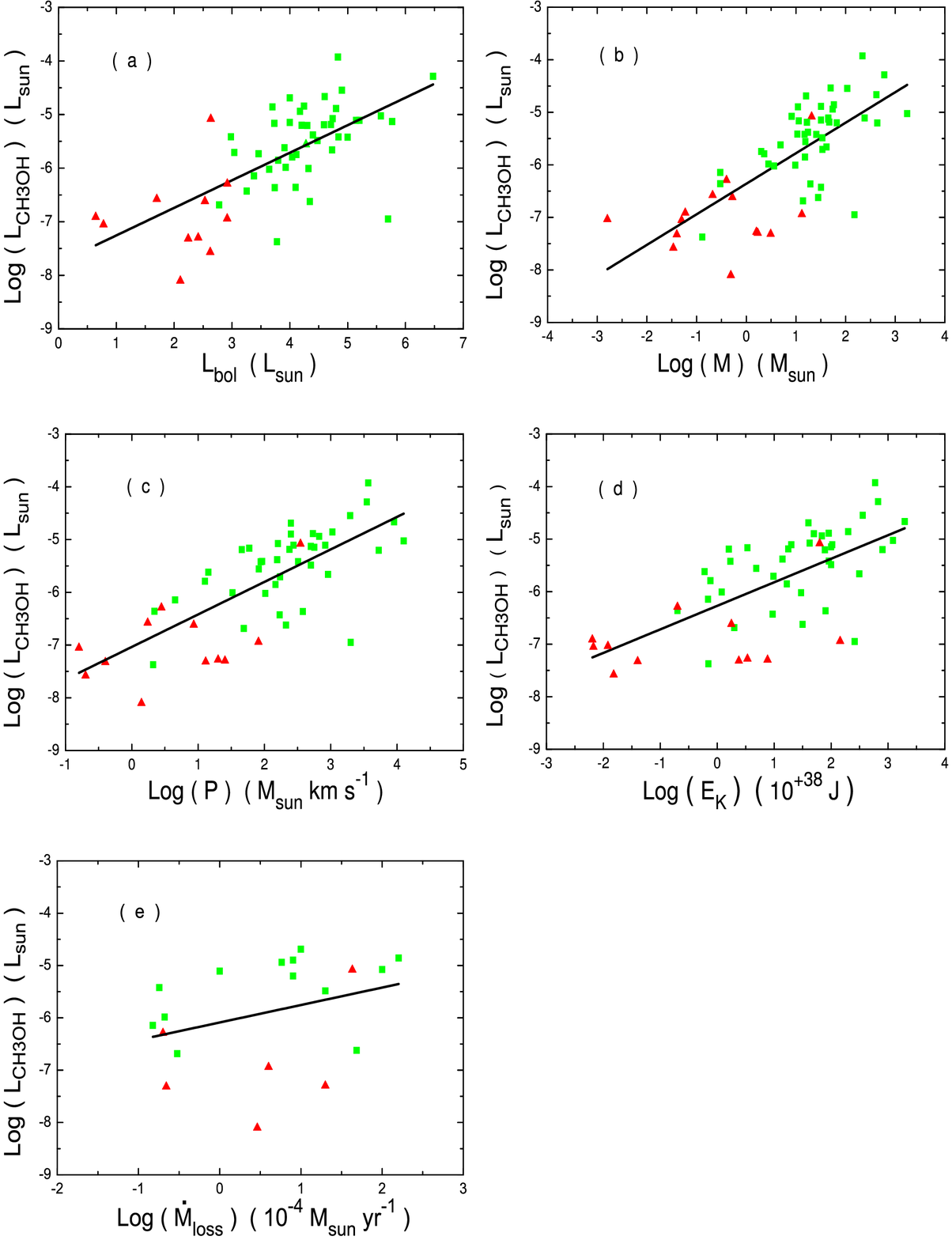,width=0.9\textwidth}
\caption{The log--log distributions between class I methanol maser luminosity
and outflow properties including bolometric luminosity of central
sources (a), outflow mass (b), momentum of outflows (c), kinetic
energy of outflows (d) and mass loss rate of central stars (e). The green squares and red triangles represent high-mass and low-mass sources, respetively. The solid black line is the best linear fit of the plot (A
color version of this figure is available in the online journal).}
\label{fig:cor}
\end{figure*}

\clearpage
\appendix
\section{Undetected 95 GHz methanol maser sources}\label{tab:undet}

\begin*
\begin{tabular}[h]{cc}

\begin{minipage}[t]{0.5\textwidth}
\centering
\begin{lrbox}{\tablebox}
\begin{tabular}{cccccc}
\hline
Name&Other name&RA&DEC&RMS&Mass\\
&&(J2000)&(J2000)&(Jy)&\\
(1)&(2)&(3)&(4)&(5)&(6)\\
\hline
G1.36+20.97 &L43&16:34:29.3&-15:47:01.0&1.62&L\\
G8.66+22.18 &16442-0930&16:46:58.4&-09:35:22.0&0.54&L\\
G11.30-2.02 &18150-2016&18:17:59.2&-20:06:58.0&1.54&H\\
G11.42-1.68 &18139-1952&18:16:56.9&-19:51:08.0&1.28&H\\
G11.42-1.37 &18128-1943&18:15:47.7&-19:42:18.0&0.74&H\\
G11.50-1.48 &18134-1942&18:16:21.7&-19:41:31.0&0.85&H\\
G17.64+0.15 &18196-1331&18:22:26.8&-13:30:15.0&1.35&H\\
G23.57+1.58 &18258-0737&18:28:34.6&-07:35:31.0&1.12&H\\
G24.89+5.38 &L483&18:17:30.2&-04:39:38.0&0.6&L\\
G28.58-1.73 &18470-0044&18:49:41.1&-04:39:35.0&0.79&H\\
G28.75+3.52 &W40&18:31:16.2&-02:06:49.0&1.43&H\\
G30.64+3.25 &18331-0035&18:35:42.4&-00:33:18.0&0.69&L\\
G31.59+5.35 &S68 Firs1&18:29:57.4&01:14:45&0.99&L\\
G31.59+5.38 &Serpens SMM1-11&18:29:50.4&01:15:19&1.32&L\\
G39.39-0.14 &19012+0536&19:03:45.4&05:40:40&1.35&H\\
G39.82-11.92 &S87&19:46:20.5&00:35:24&1.59&H\\
G40.62-0.14 &19035+0641&19:06:01.2&06:46:35&0.67&H\\
G42.16-11.49 &19471+2641&19:49:09.8&02:48:52&1.44&H\\
G42.64-10.17 &L810&19:45:24.2&03:51:01&1.38&H\\
G43.23-12.57 &19529+2704&19:54:59.7&03:12:52&0.54&H\\
G43.83-14.82 &CB214&20:03:59.1&02:38:14&1.3&L\\
G43.94-11.94 &19520+2759&19:54:06.4&04:07:25&1.08&H\\
G44.93-6.55 &B335&19:36:59.7&07:34:07&1.42&L\\
G45.12+0.13 &G45.12+0.13&19:13:28.6&10:53:22&1.33&H\\
G46.17-1.55 &AS353&19:21:30.6&11:02:14&1.41&L\\
G46.31-1.19 &L673&19:20:29.2&11:19:40&1.27&L\\
G46.34-1.15 &L673 SMM1&19:20:25.2&11:22:17&1.16&L\\
G46.53-1.02 &CB188&19:20:17.9&11:35:57&1.31&H\\
G48.57-10.19 &19550+3248&19:56:55.0&08:56:32&0.72&L\\
G50.84-11.92 &20056+3350&20:07:31.5&09:59:39&0.65&H\\
G50.92-12.07 &20062+3350&20:08:12.6&09:59:20&1.2&L\\
G51.25-27.69 &L988-a&21:02:23.0&02:03:06&0.65&L\\
G51.36-27.26 &V1331Cyg&21:01:09.0&02:21:46&0.93&L\\
G51.58-28.00 &L988-f&21:04:03.2&02:07:49&1.14&L\\
G51.67-27.92 &L988-e&21:03:58.0&02:14:38&1.22&L\\
G52.10+1.04 &19213+1723&19:23:37.1&17:28:59&1.63&H\\
G52.98+3.05 &L723 2 Flows&19:17:53.9&19:12:19&1.2&L\\
G53.16-11.98 &20106+3545&20:12:31.3&11:54:46&1.31&H\\
G53.60-10.36 &CB217&20:07:45.9&13:07:01&1.27&L\\
G53.63+0.02 &19282+1814&19:30:28.4&18:20:53&0.66&H\\
G53.78-24.68 &20582+7724&20:57:10.6&05:35:46&1.11&L\\
G53.88-15.15 &20231+3440 SMM1&20:25:00.3&10:50:05&0.7&L\\
G53.93-15.13 &20231+3440 SMM2&20:25:01.2&10:53:05&0.94&L\\
G55.72-13.06 &G75N&20:21:42.1&13:27:20&1.36&L\\
G55.86-12.99 &G75.78NE&20:21:44.8&13:37:00&1.03&H\\
G57.08-33.39 &21307+5049&21:32:31.2&03:02:22&0.94&H\\
G57.39-34.01 &21334+5039&21:35:09.2&02:53:09&0.63&H\\
G57.61-35.37 &V645Cyg&21:39:58.6&02:14:22&0.7&H\\
G58.15+3.51 &L778&19:26:32.2&23:58:42&1.36&L\\
G59.11-13.13 &AFGL2591&20:29:24.9&16:11:19&1.15&H\\
G59.13-20.12 &20520+6003&20:53:13.9&12:14:44&1.27&L\\
G59.15-20.29 &20526+5958&20:53:50.4&12:09:46&0.59&L\\
G59.24-11.32 &20216+4107&20:23:24.2&17:17:40&1.28&H\\
G59.27-13.20 &20281+4006&20:30:00.9&16:16:36&1.15&L\\
G59.36-12.88 &20272+4021&20:29:05.3&16:31:58&1.1&H\\
G59.36-0.21 &19411+2306&19:43:18.0&23:13:59&0.34&H\\
G59.47-0.05 &1548C27&19:42:55.7&23:24:20&1.23&L\\
G59.60+0.92 &19374+2352&19:39:32.8&23:59:55&1.44&H\\
G60.35-10.94 &20228+4215&20:24:34.5&18:25:01&1.19&H\\
G61.30-13.63 &20343+4129 d&20:36:07.6&17:40:01&0.85&H\\
G62.22-4.54 &CB216&20:05:54.1&23:27:04&1.57&L\\
G62.68-33.11 &21413+5442&21:43:01.2&06:56:18&0.59&H\\
G63.11-12.29 &20353+6742&20:35:45.4&19:52:59&0.56&L\\
G63.71-12.84 &L1157&20:39:06.9&20:02:13&0.77&L\\
\end{tabular}
\end{lrbox}
\scalebox{0.5}{\usebox{\tablebox}}
\end{minipage}

\begin{minipage}[t]{0.5\textwidth}
\centering
\begin{lrbox}{\tablebox}
\begin{tabular}{cccccc}
\hline
Name&Other name&RA&DEC&RMS&Mass\\
&&(J2000)&(J2000)&(Jy)&\\
(1)&(2)&(3)&(4)&(5)&(6)\\
\hline
G63.73-31.59 &GN21.38.9&21:40:29.8&08:35:13&1.17&L\\
G64.60-14.18 &Pvcep&20:45:53.6&19:57:39&1.46&L\\
G65.55-32.10 &IC1396-E&21:46:07.0&09:26:23&1.05&L\\
G66.12-34.15 &21519+5613&21:53:38.9&08:27:46&0.96&H\\
G66.76-16.43 &V1057Cyg&20:58:53.5&20:15:29&0.46&L\\
G66.95-41.06 &22142+5206&22:16:10.7&04:21:25&0.93&H\\
G67.01-17.30 &L1172D&21:02:24.0&19:54:27&0.67&L\\
G67.17-17.10 &21015+6757&21:02:09.5&20:09:09&1.54&L\\
G69.72-19.82 &CB230&21:17:40.0&20:17:32&0.73&L\\
G70.43-45.27 &22343+7501&22:35:24.3&03:17:06&1.25&L\\
G71.14-39.02 &22172+5549&22:19:09.3&08:04:45&1.15&H\\
G71.19-45.93 &22376+7455&22:38:47.3&03:11:29&1.11&L\\
G71.38-16.87 &B361&21:12:26.2&23:24:24&1.14&L\\
G71.94-35.92 &22103+5828&22:12:07.8&10:43:33&1.05&L\\
G72.26-23.87 &CB232&21:37:10.9&19:20:36&1.1&L\\
G72.34-25.73 &LKHa234&21:43:06.1&18:06:52&0.71&H\\
G72.67-36.37 &22134+5834&22:15:08.9&10:49:09&0.55&H\\
G76.20-39.62 &22305+5803&22:32:24.2&10:18:58&1.01&H\\
G76.94-22.25 &21432+4719&21:45:10.2&23:33:21&0.79&L\\
G76.98-22.12 &21429+4726&21:44:51.9&23:40:31&0.85&L\\
G77.03-22.02 &21428+4732&21:44:43.9&23:46:45&0.73&L\\
G77.12-22.35 &21441+4722&21:45:59.3&23:36:04&1.52&L\\
G77.31-22.61 &EL1-12&21:47:20.8&23:32:05&0.82&L\\
G77.49-22.66 &21461+4722&21:48:00.8&23:36:38&1.2&L\\
G77.53-33.61 &S140-N&22:19:28.6&15:32:56&0.81&L\\
G78.03-23.66 &BD+46 3471&21:52:34.6&23:13:43&1.07&L\\
G78.21-33.68 &L1204-A&22:21:27.9&15:51:42&1.3&H\\
G80.15-34.54 &L1206&22:28:52.0&16:13:43&1.35&H\\
G81.79-41.02 &22475+5939&22:49:29.6&11:54:57&0.97&H\\
G83.48-30.65 &L1221&22:28:02.7&21:01:13&1&L\\
G84.01-42.75 &22570+5912&22:59:06.7&11:28:28&1.22&H\\
G84.50-54.06 &CB244&23:25:45.8&02:17:38&1.38&L\\
G85.13-31.50 &22336+6855&22:35:06.1&21:10:53&0.68&L\\
G85.19-40.26 &Cep A&22:56:17.9&14:01:46&1.27&H\\
G89.19-40.94 &MBM 55&23:08:23.7&15:05:16&1.19&H\\
G90.23-43.99 &MWC1080&23:17:27.2&12:50:16&0.59&H\\
G90.40-43.15 &23140+6121&23:16:11.8&13:37:45&1.02&H\\
G94.25-42.40 &L1246 SMM1&23:25:04.8&15:36:40&0.58&L\\
G95.21-45.81 &23314+6033&23:33:44.6&12:50:30&0.49&H\\
G97.69-46.21 &23385+6053&23:40:51.1&13:10:29&1.13&H\\
G104.92-43.54 &23545+6508&23:57:05.1&17:25:11&1.39&H\\
G107.26-50.82 &LKHA 198&00:11:25.6&10:49:47&0.91&H\\
G110.24-45.48 &00117+6412&00:14:26.7&16:28:44&0.48&H\\
G113.80-44.60 &00213+6530&00:24:10.4&17:47:02&0.98&L\\
G115.28-45.09 &00259+6510&00:28:49.1&17:26:47&0.91&L\\
G117.95-46.66 &00342+6374&00:37:13.0&16:04:15&1.21&H\\
G131.56-45.61 &01133+6434&01:16:37.2&16:50:39&1.24&H\\
G154.86-42.86 &W3-IRS5&02:25:40.5&14:05:52&1.29&H\\
G156.19-42.92 &IC1805-W&02:29:02.3&13:33:32&1.16&H\\
G157.66-42.02 &02310+6133&02:34:46.4&13:46:22&1.44&H\\
G161.63-39.79 &02461+6147&02:50:09.5&13:59:58&0.97&H\\
G168.75-39.96 &AFGL437&03:07:24.4&10:31:08&0.74&H\\
G173.06-36.36 &AFGL490-iki&03:27:28.0&10:54:10&1.65&H\\
G173.20-36.41 &AFGL490&03:27:38.6&10:47:04&1.5&H\\
G176.26-39.68 &RNO13&03:25:09.5&06:46:22&1.23&L\\
G176.37-39.62 &L1448&03:25:36.5&06:45:19&1.02&L\\
G176.40-39.62 &L1448 U-star&03:25:38.5&06:44:04&1.13&L\\
G176.61-38.74 &NGC1333&03:28:39.5&07:13:34&0.75&L\\
G176.64-38.69 &IRAS2NNE-SSW&03:28:53.5&07:14:53&0.71&L\\
G176.65-38.60 &HH6&03:29:10.8&07:18:19&1.13&L\\
G176.66-38.64 &HH7-11 SSV&03:29:03.7&07:16:04&1.4&L\\
G177.24-39.63 &L1455NW&03:27:26.1&06:15:48&1.26&L\\
G177.34-39.62 &RNO15FIR&03:27:40.1&06:13:03&0.74&L\\
G177.38-39.60 &L1455M&03:27:48.1&06:12:06&1.67&L\\
\end{tabular}
\end{lrbox}
\scalebox{0.5}{\usebox{\tablebox}}
\end{minipage}
\end{tabular}
\end*

\begin{table}
\begin{tabular}{cc}
\begin{minipage}[t]{0.5\textwidth}
\centering
\begin{lrbox}{\tablebox}
\begin{tabular}{cccccc}
\hline
Name&Other name&RA&DEC&RMS&Mass\\
&&(J2000)&(J2000)&(Jy)&\\
(1)&(2)&(3)&(4)&(5)&(6)\\
\hline
G177.63-38.58 &03282+3035&03:31:20.2&06:45:25&1.32&L\\
G177.73-37.96 &03301+3057&03:33:22.8&07:07:30&0.7&L\\
G177.73-39.03 &03271+3013&03:30:14.6&06:23:49&0.84&L\\
G178.78-26.46 &PP 13S&04:10:41.1&14:07:54&0.64&L\\
G178.86-19.97 &HL/XZ Tau&04:31:38.0&18:13:59&1.41&L\\
G178.92-34.02 &B5-IRS4&03:47:45.9&09:03:45&1.17&L\\
G178.93-20.05 &L1551-IRS5&04:31:33.9&18:08:05&0.69&L\\
G178.93-20.03 &L1551NE&04:31:36.9&18:08:35&1.39&L\\
G179.08-34.17 &B5-IRS1&03:47:41.6&08:51:43&0.7&L\\
G179.09-34.38 &B5-IRS3&03:47:05.4&08:43:09&1.41&L\\
G179.11-35.42 &HH211&03:43:57.1&08:00:50&1.29&L\\
G179.19-34.13 &B5-IRS2&03:48:03.6&08:49:28&1.3&L\\
G179.56-23.49 &04191+1523&04:21:59.5&15:30:17&1.29&L\\
G186.95-3.84 &CB34&05:47:02.3&21:00:10&1&L\\
G187.96-12.83 &05137+3919&05:17:13.5&15:22:23&0.85&H\\
G188.51-34.95 &L1489&04:04:43.5&02:18:57&1.21&L\\
G189.03+0.78 &AFGL 6366S&06:08:41.1&21:31:01&1.01&H\\
G190.20-31.37 &04166+2706&04:19:42.7&03:13:40&1.21&L\\
G190.63-31.18 &04181+2655&04:21:10.5&03:02:06&0.69&L\\
G190.75-13.70 &05168+3634&05:20:16.5&12:37:21&1.72&H\\
G192.07-11.00 &RNO43N&05:32:27.0&12:57:06&1.44&L\\
G192.13-11.06 &RNO43S&05:32:23.9&12:52:07&1.23&L\\
G192.16-11.11 &L1582B&05:32:15.8&12:49:20&1.09&L\\
G192.16-3.82 &05553+1631&05:58:13.6&16:32:00&0.63&H\\
G192.16-11.08 &RNO43&05:32:21.8&12:49:40&0.8&L\\
G192.88-3.18 &HD250550&06:02:00.2&16:13:04&1.05&L\\
G192.99+0.15 &06114+1745&06:14:24.1&17:44:36&0.96&H\\
G193.81-31.25 &04239+2436&04:26:56.9&00:43:36&1.16&L\\
G193.86-10.32 &05358+3543 A+B&05:38:34.6&11:47:35&1.01&H\\
G194.37-30.83 &L1524&04:29:23.8&00:32:58&1&L\\
G194.45-30.44 &ZZ Tau&04:30:53.0&00:41:50&1.08&L\\
G194.52-27.82 &L1527&04:39:53.3&02:03:06&1.19&L\\
G194.58-28.04 &04361+2547&04:39:14.0&01:53:22&1.34&L\\
G194.81-30.30 &TMC2A&04:31:59.8&00:30:49&1.1&L\\
G194.82-28.07 &TMC1A&04:39:34.8&01:41:46&1.62&L\\
G194.82-27.96 &IC2087&04:39:58.9&01:45:06&1.88&L\\
G194.99-27.68 &04381+2540&04:41:13.0&01:46:37&0.74&L\\
G195.03-11.71 &05327+3404&05:36:05.7&10:06:12&1.26&L\\
G195.05-30.21 &L1529&04:32:44.7&00:23:13&0.88&L\\
G195.27-16.98 &HH114/115&05:18:17.0&07:11:01&1.44&H\\
G195.72-29.75 &04325+2402&04:35:33.5&00:08:15&1.34&L\\
G196.93-10.42 &B35&05:44:26.4&09:08:14&0.91&L\\
G197.11-12.41 &AFGL5157&05:37:48.2&07:59:24&0.7&H\\
G198.58-9.14 &L1598NW&05:52:11.5&08:21:28&1.16&L\\
G201.87-7.62 &S241&06:03:53.7&06:14:44&0.84&H\\
G202.12+2.65 &Mon OB1I&06:41:05.4&10:49:46&1.07&L\\
G202.30+2.54 &Mon OB1H&06:41:03.1&10:37:03&0.97&L\\
G202.93+2.26 &Mon OB1G&06:41:12.3&09:55:35&1.78&L\\
G203.23+2.07 &Mon OB1D&06:41:03.9&09:34:39&1.1&L\\
\hline
\end{tabular}
\end{lrbox}
\scalebox{0.5}{\usebox{\tablebox}}
\end{minipage}

\begin{minipage}[t]{0.5\textwidth}
\centering
\begin{lrbox}{\tablebox}
\begin{tabular}{cccccc}
\hline
Name&Other name&RA&DEC&RMS&Mass\\
&&(J2000)&(J2000)&(Jy)&\\
(1)&(2)&(3)&(4)&(5)&(6)\\
\hline
G203.32-11.94 &05487+0255&05:51:23.1&02:55:45&1.03&L\\
G203.36-11.73 &05490+2658&05:52:12.7&02:59:33&1.08&H\\
G203.47-11.92 &05491+0247&05:51:46.0&02:48:35&1.26&L\\
G203.76+1.27 &NGC2261&06:39:09.9&08:44:12&1.02&L\\
G204.88-13.85 &NGC2071N&05:47:34.5&00:41:00&2.75&L\\
G205.38-14.41 &NGC2068/LBS17&05:46:29.7&-00:00:37.0&1.02&L\\
G205.42-14.42 &NGC2068&05:46:31.7&-00:02:56.0&1.18&L\\
G205.52-14.57 &HH24&05:46:11.5&-00:12:17.0&0.93&L\\
G205.95-17.09 &Ori-I-2&05:38:04.7&-01:45:09.0&1.36&L\\
G206.01-15.48 &HH212&05:43:51.5&-01:02:52.0&0.61&L\\
G206.56-16.34 &NGC2024 Ori B&05:41:49.5&-01:55:17.0&0.83&H\\
G206.84-2.38 &06291+0421&06:31:47.8&04:19:31&0.93&H\\
G206.86-16.55 &NGC 2023&05:41:37.1&-02:15:58.0&1.19&L\\
G206.86-16.61 &NGC2023-MM1&05:41:25.0&-02:18:09.0&1.51&L\\
G207.33-2.15 &06308+0402&06:33:31.4&04:00:07&1.05&H\\
G207.60-23.03 &L1634&05:19:49.0&-05:52:05.0&1.33&L\\
G208.63-19.21 &CSO 2&05:35:13.9&-04:59:22.0&1.48&L\\
G208.66-19.21 &AC 3&05:35:18.3&-05:00:41.0&1.06&L\\
G208.75-19.22 &OMC-2/3(MMS 9)&05:35:25.8&-05:05:37.0&0.85&L\\
G208.77-19.19 &MMS 10&05:35:33.8&-05:05:38.0&2.02&L\\
G208.90-20.05 &Ori A-W&05:32:42.2&-05:35:48.0&1.56&L\\
G209.25-19.12 &05341-0539&05:36:38.4&-05:28:16.0&1.46&L\\
G209.85-20.27 &HH83&05:33:32.2&-06:29:44.0&1.03&L\\
G210.04-19.81 &HH 34&05:35:30.2&-06:26:57.0&0.72&L\\
G210.35-19.69 &V380/OriNE&05:36:26.0&-06:39:12.0&1.36&L\\
G210.40-19.72 &V380 Ori&05:36:25.9&-06:42:38.0&1.3&L\\
G210.43-19.74 &05339-0646&05:36:23.9&-06:44:45.0&0.88&L\\
G210.44-19.76 &CS-star HH1&05:36:20.8&-06:45:35.0&1.55&L\\
G210.44-19.75 &VLA3&05:36:22.9&-06:45:22.0&1.55&L\\
G210.45-19.76 &HH 1-2&05:36:22.8&-06:46:07.0&1.96&L\\
G210.58-19.81 &V380/OriS&05:36:25.7&-06:54:12.0&1.49&L\\
G210.82-36.61 &L1642&04:34:49.9&-14:13:09.0&0.94&L\\
G210.96-19.34 &05363-0702&05:38:44.5&-07:01:03.0&1.04&L\\
G211.44-19.39 &Haro 4-255&05:39:22.0&-07:26:45.0&0.99&L\\
G211.57-19.29 &L1641S3&05:39:56.0&-07:30:26.0&0.45&L\\
G211.58-19.15 &L1641S&05:40:28.0&-07:27:28.0&0.99&L\\
G212.25-19.36 &L1641S4&05:40:49.2&-08:06:51.0&1.36&L\\
G212.63-19.00 &L1641S2&05:42:47.0&-08:17:06.0&0.98&L\\
G213.88-11.83 &GGD 12-15&06:10:51.5&-06:11:27.0&2.52&H\\
G217.30-0.05 &BFS 56&06:59:14.4&-03:54:52.0&0.53&L\\
G217.38-0.08 &BiP 14&06:59:16.3&-03:59:39.0&0.62&H\\
G218.02-0.32 &S287-C&06:59:36.5&-04:40:22.0&1.72&L\\
G218.06-0.11 &S287-A&07:00:23.6&-04:36:38.0&0.54&L\\
G218.10-0.37 &S287-B&06:59:34.4&-04:46:00.0&1.77&L\\
G218.15-15.00 &06047-1117&06:06:41.4&-11:18:40.0&1.05&L\\
G224.35-2.01 &07028 1100&07:05:13.2&-11:04:41.0&0.73&L\\
G224.61-2.56 &Z Cma&07:03:43.6&-11:33:06.0&1.02&L\\
G228.99-4.62 &CB54&07:04:20.9&-16:23:20.0&0.9&L\\
G356.09+20.75 &16191 1936&16:22:04.4&-19:43:26.0&1.57&L\\
\hline
\end{tabular}
\end{lrbox}
\scalebox{0.5}{\usebox{\tablebox}}
\end{minipage}
\end{tabular}
\begin{tablenotes}
\item Columns (1)--(6) list the source name, other source name from \citet[]{wu04} catalog, corresponding equatorial coordinates, rms noise and mass type (L represents low-mass sources, H represents high-mass sources), respectively
\end{tablenotes}
\end{table}


\begin{thebibliography}{}

\expandafter\ifx\csname natexlab\endcsname\relax\def\natexlab#1{#1}\fi

\bibitem[Arce et al.(2007)]{arc07} Arce, H.~G., Shepherd, D.,
Gueth, F., et al.\ 2007, Protostars and Planets V, 245

\bibitem[Bae et al.(2011)]{bae11} Bae, J.-H., Kim, K.-T.,
Youn, S.-Y., et al.\ 2011, \apjs, 196, 21

\bibitem[Bartkiewicz
\& van Langevelde(2012)]{bar12} Bartkiewicz, A., \& van Langevelde, H.~J.\ 2012, IAU Symposium, 287, 117

\bibitem[Batrla et al.(1987)]{bat87} Batrla, W., Matthews,
H.~E., Menten, K.~M., \& Walmsley, C.~M.\ 1987, \nat, 326, 49


\bibitem[Breen et al.(2007)]{bre07} Breen, S.~L., Ellingsen,
S.~P., Johnston-Hollitt, M., et al.\ 2007, \mnras, 377, 491


\bibitem[Breen et al.(2011)]{bre11} Breen, S.~L., Ellingsen,
S.~P., Caswell, J.~L., et al.\ 2011, \apj, 733, 80

\bibitem[Caswell(2009)]{cas09} Caswell, J.~L.\ 2009, \pasa,
26, 454

\bibitem[Caswell et al.(2010)]{cas10} Caswell, J.~L., Fuller,
G.~A., Green, J.~A., et al.\ 2010, \mnras, 404, 1029

\bibitem[Caswell et al.(2011)]{cas11} Caswell, J.~L., Fuller,
G.~A., Green, J.~A., et al.\ 2011, \mnras, 417, 1964

\bibitem[Chambers et al.(2011)]{cha11} Chambers, E.~T.,
Yusef-Zadeh, F., \& Roberts, D.\ 2011, \apj, 733, 42

\bibitem[Chen et al.(2009)]{che09} Chen, X., Ellingsen,
S.~P., \& Shen, Z.-Q.\ 2009, \mnras, 396, 1603

\bibitem[Chen et al.(2011)]{che11} Chen, X., Ellingsen,
S.~P., Shen, Z.-Q., Titmarsh, A., \& Gan, C.-G.\ 2011, \apjs, 196, 9

\bibitem[Chen et al.(2012)]{che12} Chen, X., Ellingsen,
S.~P., He, J.-H., et al.\ 2012, \apjs, 200, 5

\bibitem[Cragg et al.(1992)]{cra92} Cragg, D.~M., Johns,
K.~P., Godfrey, P.~D., \& Brown, R.~D.\ 1992, \mnras, 259, 203

\bibitem[Cragg et al.(2005)]{cra05} Cragg, D.~M., Sobolev,
A.~M., \& Godfrey, P.~D.\ 2005, \mnras, 360, 533

\bibitem[Cyganowski et al.(2008)]{cyg08} Cyganowski, C.~J.,
Whitney, B.~A., Holden, E., et al.\ 2008, \aj, 136, 2391

\bibitem[Cyganowski et al.(2009)]{cyg09} Cyganowski, C.~J.,
Brogan, C.~L., Hunter, T.~R., \& Churchwell, E.\ 2009, \apj, 702, 1615

\bibitem[Ellingsen(2005)]{ell05} Ellingsen, S.~P.\ 2005,
\mnras, 359, 1498

\bibitem[Ellingsen(2006)]{ell06} Ellingsen, S.~P.\ 2006,
\apj, 638, 241

\bibitem[Ellingsen et al.(2007)]{ell07} Ellingsen, S.~P.,
Voronkov, M.~A., Cragg, D.~M., et al.\ 2007, IAU Symposium, 242, 213

\bibitem[Ellingsen et al.(2011)]{ell11} Ellingsen, S.~P.,
Breen, S.~L., Sobolev, A.~M., et al.\ 2011, \apj, 742, 109

\bibitem[Fish et al.(2011)]{fis11} Fish, V.~L., Muehlbrad,
T.~C., Pratap, P., et al.\ 2011, \apj, 729, 14

\bibitem[Fontani et al.(2010)]{fon10} Fontani, F., Cesaroni, R., \& Furuya, R.~S.\ 2010, \aap, 517, A56

\bibitem[Garay et al.(2002)]{gar02} Garay, G., Mardones, D.,
Rodr{\'{\i}}guez, L.~F., Caselli, P.,
\& Bourke, T.~L.\ 2002, \apj, 567, 980

\bibitem[Gibb
\& Davis(1998)]{gib98} Gibb, A.~G., \& Davis, C.~J.\ 1998, \mnras, 298, 644

\bibitem[Green et al.(2009)]{gre09} Green, J.~A., Caswell,
J.~L., Fuller, G.~A., et al.\ 2009, \mnras, 392, 783

\bibitem[Green et al.(2010)]{gre10} Green, J.~A., Caswell,
J.~L., Fuller, G.~A., et al.\ 2010, \mnras, 409, 913

\bibitem[Green et al.(2012)]{gre12} Green, J.~A., Caswell,
J.~L., Fuller, G.~A., et al.\ 2012, \mnras, 420, 3108

\bibitem[Haschick et al.(1990)]{has90} Haschick, A.~D.,
Menten, K.~M., \& Baan, W.~A.\ 1990, \apj, 354, 556

\bibitem[Johnston et al.(1997)]{joh97} Johnston, K.~J.,
Gaume, R.~A., Wilson, T.~L., Nguyen, H.~A.,
\& Nedoluha, G.~E.\ 1997, \apj, 490, 758

\bibitem[Kalenskii et
al.(1994)]{kal94} Kalenskii, S.~V., Liljestroem, T., Val'tts, I.~E., et al.\ 1994, \aaps, 103, 129

\bibitem[Kalenskii et al.(2001)]{kal01} Kalenskii,
S.~V., Slysh, V.~I., Val'tts, I.~E., Winnberg, A.,
\& Johansson, L.~E.\ 2001, Astronomy Reports, 45, 26

\bibitem[Kalenskii et al.(2006)]{kal06} Kalenskii,
S.~V., Promyslov, V.~G., Slysh, V.~I., Bergman, P.,
\& Winnberg, A.\ 2006, Astronomy Reports, 50, 289

\bibitem[Kalenskii et al.(2010)]{kal10} Kalenskii, S.~V.,
Johansson, L.~E.~B., Bergman, P., Kurtz, S., Hofner, P., Walmsley, C.~M.,
\& Slysh, V.~I.\ 2010, \mnras, 405, 613

\bibitem[Kurtz et al.(2004)]{kur04} Kurtz, S., Hofner, P.,
\& {\'A}lvarez, C.~V.\ 2004, \apjs, 155, 149

\bibitem[Larionov et
al.(1999)]{lar99} Larionov, G.~M., Val'tts, I.~E., Winnberg, A., et al.\ 1999, \aaps, 139, 257

\bibitem[Larionov
\& Val'tts(2007)]{lar07} Larionov, G.~M., \& Val'tts, I.~E.\ 2007, Astronomy Reports, 51, 756


\bibitem[Leurini et
al.(2004)]{leu04} Leurini, S., Schilke, P., Menten, K.~M., et al.\ 2004, \aap, 422, 573

\bibitem[Leurini et
al.(2007)]{leu07} Leurini, S., Schilke, P., Wyrowski, F., \& Menten, K.~M.\ 2007, \aap, 466, 215

\bibitem[Liechti
\& Wilson(1996)]{lie96} Liechti, S., \& Wilson, T.~L.\ 1996, \aap, 314, 615


\bibitem[Litovchenko et al.(2011)]{lit11} Litovchenko, I.~D.,
Alakoz, A.~V., Val'Tts, I.~E.,
\& Larionov, G.~M.\ 2011, Astronomy Reports, 55, 1086

\bibitem[Menten(1991)]{men91} Menten, K.~M.\ 1991, \apjl,
380, L75

\bibitem[Minier et
al.(2003)]{min03} Minier, V., Ellingsen, S.~P., Norris, R.~P., \& Booth, R.~S.\ 2003, \aap, 403, 1095

\bibitem[Pandian et al.(2007)]{pan07} Pandian, J.~D.,
Goldsmith, P.~F., \& Deshpande, A.~A.\ 2007, \apj, 656, 255

\bibitem[Pestalozzi et
al.(2005)]{pes05} Pestalozzi, M.~R., Minier, V., \& Booth, R.~S.\ 2005, \aap, 432, 737


\bibitem[Pihlstr{\"o}m et al.(2011)]{pih11} Pihlstr{\"o}m,
Y.~M., Sjouwerman, L.~O., \& Fish, V.~L.\ 2011, \apjl, 739,
L21

\bibitem[Plambeck
\& Menten(1990)]{pla90} Plambeck, R.~L., \& Menten, K.~M.\ 1990, \apj, 364, 555

\bibitem[Purcell et al.(2009)]{pur09} Purcell, C.~R.,
Longmore, S.~N., Burton, M.~G., et al.\ 2009, \mnras, 394, 323

\bibitem[Rygl et
al.(2010)]{ry10} Rygl, K.~L.~J., Brunthaler, A., Reid, M.~J., et al.\ 2010, \aap, 511, A2

\bibitem[Sandell et al.(2003)]{san03} Sandell, G., Wright,
M., \& Forster, J.~R.\ 2003, \apjl, 590, L45

\bibitem[Sandell et al.(2005)]{san05} Sandell, G., Goss,
W.~M., \& Wright, M.\ 2005, \apj, 621, 839

\bibitem[Sarma
\& Momjian(2009)]{sar09} Sarma, A.~P., \& Momjian, E.\
2009, \apjl, 705, L176

\bibitem[Sarma
\& Momjian(2011)]{sar11} Sarma, A.~P., \& Momjian, E.\
2011, \apjl, 730, L5

\bibitem[Slysh et al.(1994)]{sly94} Slysh, V.~I., Kalenskii,
S.~V., Valtts, I.~E., \& Otrupcek, R.\ 1994, \mnras, 268, 464

\bibitem[Sobolev et al.(2005)]{sob05} Sobolev, A.~M.,
Ostrovskii, A.~B., Kirsanova, M.~S., et al.\ 2005, Massive Star Birth: A
Crossroads of Astrophysics, 227, 174


\bibitem[Val'tts et al.(1995)]{val95} Val'tts, I.~E., Dzyura,
A.~M., Kalenskii, S.~V., et al.\ 1995, \azh, 72, 22

\bibitem[Val'tts et al.(2000)]{val00} Val'tts, I.~E.,
Ellingsen, S.~P., Slysh, V.~I., Kalenskii, S.~V., Otrupcek, R.,
\& Larionov, G.~M.\ 2000, \mnras, 317, 315

\bibitem[Val'tts
\& Larionov(2007)]{val07} Val'tts, I.~E., \& Larionov, G.~M.\ 2007, Astronomy Reports, 51, 519

\bibitem[Voronkov(1999)]{vor99} Voronkov, M.~A.\ 1999,
Astronomy Letters, 25, 149

\bibitem[Voronkov et al.(2005)]{vor05} Voronkov, M.~A.,
Sobolev, A.~M., Ellingsen, S.~P.,
\& Ostrovskii, A.~B.\ 2005, \mnras, 362, 995

\bibitem[Voronkov et al.(2006)]{vor06} Voronkov, M.~A.,
Brooks, K.~J., Sobolev, A.~M., et al.\ 2006, \mnras, 373, 411

\bibitem[Voronkov et al.(2010a)]{vor10a} Voronkov, M.~A.,
Caswell, J.~L., Ellingsen, S.~P.,
\& Sobolev, A.~M.\ 2010a, \mnras, 405, 2471

\bibitem[Voronkov et al.(2010b)]{vor10b} Voronkov, M.~A.,
Caswell, J.~L., Britton, T.~R., Green, J.~A., Sobolev, A.~M.,
\& Ellingsen, S.~P.\ 2010b, \mnras, 408, 133

\bibitem[Voronkov et al.(2011)]{vor11} Voronkov, M.~A.,
Walsh, A.~J., Caswell, J.~L., et al.\ 2011, \mnras, 413, 2339

\bibitem[Voronkov et al.(2012)]{vor12} Voronkov, M.~A.,
Caswell, J.~L., Ellingsen, S.~P., et al.\ 2012, IAU Symposium, 287, 433

\bibitem[Wirstr{\"o}m et
al.(2011)]{wir11} Wirstr{\"o}m, E.~S., Geppert, W.~D., Hjalmarson, {\AA}., et al.\ 2011, \aap, 533, A24

\bibitem[Wu et
al.(2004)]{wu04} Wu, Y., Wei, Y., Zhao, M., Shi, Y., Yu, W., Qin, S., \& Huang, M.\ 2004, \aap, 426, 503

\bibitem[Xu et al.(2006)]{xu06} Xu, Y., Reid, M.~J., Zheng,
X.~W., \& Menten, K.~M.\ 2006, Science, 311, 54

\bibitem[Xu et
al.(2008)]{xu08} Xu, Y., Li, J.~J., Hachisuka, K., et al.\ 2008, \aap, 485, 729

\bibitem[Xu et
al.(2009)]{xu09} Xu, Y., Voronkov, M.~A., Pandian, J.~D., et al.\ 2009, \aap, 507, 1117

\end{thebibliography}
\end{document}